\documentclass[letterpaper]{JHEP3}

\usepackage{amsmath}
\usepackage{amssymb}
\usepackage{amsfonts}
\usepackage{graphicx}
\usepackage{bm}
\usepackage[hang,flushmargin]{footmisc}

\allowdisplaybreaks[4]

\newcommand{\bea}{\begin{eqnarray}}
\newcommand{\eea}{\end{eqnarray}}
\newcommand{\beq}{\begin{equation}}
\newcommand{\eeq}{\end{equation}}
\newcommand{\nn}{\nonumber}
\newcommand{\ignore}[1]{}

\def\<{\langle}
\def\>{\rangle}

\def\Hp{H_{\rm p}}
\def\Hinf{H_{\rm inf}}
\def\Hb{H_{\rm b}}

\def\oa{\overline{a}}
\def\tz{\tilde{z}}
\def\xc{x_{\rm c}}
\def\tc{\theta_{\rm c}}

\def\cH{{\cal H}}
\def\bfx{\bm{x}}

\title{Gravity waves from cosmic bubble collisions}

\author{Michael P.~Salem, Prashant Saraswat, and Edgar Shaghoulian\\
Stanford Institute for Theoretical Physics and Department of Physics, Stanford University,\\ 
Stanford, California 94305, USA}

\abstract{Our local Hubble volume might be contained within a bubble that nucleated in a false vacuum with only two large spatial dimensions.  We study bubble collisions in this scenario and find that they generate gravity waves, which are made possible in this context by the reduced symmetry of the global geometry.  These gravity waves would produce $B$-mode polarization in the cosmic microwave background, which could in principle dominate over the inflationary background.}

\begin{document}

\section{Introduction}
\label{sec:introduction}

The configuration space of string compactifications appears to contain an enormous number of local minima, known as the landscape of string vacua \cite{Bousso:2000xa,Douglas:2003um,Kachru:2003aw,Susskind:2003kw}.  According to the classical equations of motion, if a sufficiently large volume is in a state sufficiently close to a positive-energy vacuum, it will expand without bound, the local geometry rapidly approaching a patch of de Sitter space.  Meanwhile, in the context of quantum theory, transitions between vacua should occur.  A semi-classical analysis indicates that these transitions take place locally, by way of bubble formation \cite{Banks:1973ps,Coleman:1977py,Callan:1977pt,Coleman:1980aw,Brown:2007sd}.  Viewed from the outside, a given bubble nucleates with some finite initial radius, which then expands at a rate that rapidly approaches the speed of light.  Viewed from the inside, the bubble has an open Friedmann--Robertson--Walker (FRW) geometry, up to the effects of collisions with other bubbles.  Setting aside the effects of bubble collisions, the FRW symmetry of the bubble establishes suitable initial conditions for slow-roll inflation.  Thus, our local Hubble volume could be contained within one of these bubbles \cite{Gott:1982zf}.  (A bubble like ours must feature a period of slow-roll inflation to redshift away the initial spatial curvature in the bubble, followed by reheating into appropriate degrees of freedom to initiate the standard big-bang evolution.)  Including the effects of bubble collisions, the global FRW symmetry of the bubble is broken, potentially giving the opportunity to confirm this picture of cosmology by observing any associated late-time phenomena.

Taking ourselves to reside within such a bubble, one quantity of interest is the expected number of bubble collisions for which the boundary of the causal future of the collision in\-ter\-sects the surface of last scattering.  These ``partial-sky'' collisions have the potential to affect the cosmic microwave background (CMB), but their causal futures do not cover the entire CMB, a cir\-cum\-stance in which their effects might be hard to distinguish from background.  The ex\-pect\-ed number of these collisions is of order \cite{Garriga:2006hw,Aguirre:2007an,Dahlen:2008rd,Freivogel:2009it,Salem:2011qz}
\beq
{\cal N}_{\rm bub}\sim \frac{\Hp^2}{\Hinf^2}\,\frac{\Gamma}{\Hp^4}\,
\sqrt{\Omega_k}\sim \frac{\Hp^2}{\Hinf^2}\,e^{-\Delta S-\Delta N}\,,
\label{N1}
\eeq 
where $\Hinf$ is the Hubble rate during slow-roll inflation in our bubble, $\Hp$ is the Hubble rate of the ``parent'' vacuum in which our bubble nucleates, $\Gamma$ is the decay rate per unit volume in the parent vacuum, and $\Omega_k$ is the present-day curvature parameter.  In the second expression we have substituted $\Gamma/\Hp^4\sim e^{-\Delta S}$ (introducing an admittedly crude estimate of the prefactor), where $\Delta S$ is the difference between the Euclidean action of the dominant de\-cay channel and that of the unperturbed parent vacuum, and $\Omega_k\sim e^{-2\Delta N}$, where $\Delta N$ is the difference between the total number of $e$-folds of slow-roll inflation in the bubble and the number required to solve the flatness problem.

If ${\cal N}_{\rm bub}\ll 1$, then there is little hope to observe any effects from these collisions.  With respect to this, one concern is the size of $\Delta N$, which could in principle be very large.  Yet, at least according to initial attempts to model the distribution of inflationary potentials in the landscape \cite{Freivogel:2005vv,BlancoPillado:2012cb,Guth:2012ww,Yang:2012jf}, it is not unlikely---the probability is of order ten percent---that $\Delta N$ is less than four or so \cite{Bousso:2009gx,DeSimone:2009dq}.  Another concern is the size of $\Delta S$, which could also be very large, since the units of the gravitational contributions to $\Delta S$ are set by the Planck scale.  Nevertheless, in the context of a string landscape characterized by Planck-scale dynamics, it seems plausible that $\Delta S$ is not too far from order unity, though this possibility does stretch the validity of the semi-classical methods used to obtain the above result.  Some leeway with respect to the sizes of $\Delta N$ and $\Delta S$ is afforded by the factor of $\Hp^2/\Hinf^2$, which could be very large.  For example, if we assume $\Hp$ is of order the Planck scale, then the current observational bound on primordial gravity waves implies that $\Hp^2/\Hinf^2\gtrsim e^{23}$ \cite{Komatsu:2010fb}, while TeV-scale inflation in our bubble would correspond to $\Hp^2/\Hinf^2\sim e^{146}$.  Collectively, these considerations give hope that ${\cal N}_{\rm bub}\gtrsim {\cal O}(1)$.

Granting that these partial-sky collisions exist, our interest turns to their consequences for cosmological observables.  One important feature of a two-bubble collision is that it preserves too much symmetry to act as a source for gravity waves \cite{Kosowsky:1991ua,Chang:2007eq}.  Nevertheless, if the inflaton is the field composing the bubble wall, or if it is coupled to that field, then bubble collisions could generate local perturbations in the initial value of the inflaton.  Indeed, a bubble collision could cause a large disruption in its wake, if for example it creates a domain wall that rapidly accelerates into our bubble.  However, since we observe the FRW symmetry of our local spacetime to be only slightly broken, we can ignore this possibility.  Therefore, we assume the bubble collisions of interest to observational cosmology cause only small perturbations.  In this context each partial-sky collision produces a spatially localized, axisymmetric perturbation on the CMB sky, with the probability distribution for these perturbations being essentially uniform with respect to both the position on the sky and the cosine of the angular size of the perturbation \cite{Chang:2007eq,Aguirre:2007wm,Aguirre:2008wy,Chang:2008gj,Aguirre:2009ug,Czech:2010rg,Kleban:2011yc,Gobbetti:2012yq}.  (For another possible effect, see \cite{Larjo:2009mt}; for a review, see \cite{Kleban:2011pg}.)  The amplitude of this signal decreases with $\Delta N$, however it might be possible to detect these effects with $\Delta N$ as large as ten or so \cite{Chang:2007eq}.  On the other hand, since the effects are transmitted via the inflaton, they correspond to scalar metric perturbations, and thus our ability to discern them competes with cosmic variance. 

The discussion so far has assumed that all of the relevant vacua have three large spatial dimensions.  Yet, the string landscape includes compactifications with fewer and more large dimensions, and these vacua should also play a part in cosmology.  While string compactifications can be complicated, the dynamics from the perspective of a low-energy effective theory can be explored using simpler models of compactification, as in \cite{Douglas:2006es,BlancoPillado:2009di,Carroll:2009dn,BlancoPillado:2009mi,Yang:2009wz,BlancoPillado:2010uw,Adamek:2010sg,Brown:2010mg,Brown:2010mf}.  We here focus on the case where the parent vacuum has only two large spatial dimensions.  In particular, we take the size of one (globally closed) spatial dimension to be determined by a metastable modulus $\psi$:  the dimension is ``compact'' in the parent vacuum, where $\psi$ sits at a local minimum, but this state is unstable to bubble formation, during which $\psi$ tunnels out of the local minimum.  Inside the bubble, $\psi$ evolves toward infinity, in such a way that late-time observers see three large spatial dimensions with (approximate) FRW symmetry.

The reduced symmetry of the global bubble geometry in this scenario gives rise to distinctive phenomena, including for example statistical anisotropy among the inflationary perturbations in the CMB \cite{BlancoPillado:2010uw,Graham:2010hh}.  However, it turns out that this and related effects are strongly constrained by observation.  In particular, the statistical anisotropy is suppressed relative to the usual, statistically isotropic inflationary spectrum by a factor of order $\Omega_k$.  Moreover, the global anisotropy of the geometry induces a CMB quadrupole, which (absent fine-tuning) is also of order $\Omega_k$ \cite{Graham:2010hh}.  In light of the measured quadrupole, this constrains $\Omega_k$ to be of order $10^{-5}$ or less, making future detection of the statistical anisotropy seem out of reach.  Other effects of the global anisotropy of the bubble geometry are similarly constrained.

On the other hand, the effects of bubble collisions are potentially discernible even with such small values of $\Omega_k$.  In this context, the effects of bubble collisions need not (and indeed do not) correspond to axisymmetric perturbations in the sky---the corresponding rotational symmetry being broken into the product of two parity symmetries---and the centers of these perturbations always appear along a great circle on the sky \cite{Salem:2010mi}.  These provide distinctive signatures of this scenario, especially in the event that three or more bubble collisions are observed.  The distribution of the cosine of the angular sizes of these perturbations is uniform.  Meanwhile, the analogue of (\ref{N1}) is \cite{Salem:2010mi}   
\beq
{\cal N}_{\rm bub}\sim \frac{\Hp}{\Hinf}\,\frac{\Gamma}{\Hp^3}\,\sqrt{\Omega_k}\,,
\label{N2}
\eeq 
where $\Gamma$ is now the decay rate per unit two-dimensional volume in the (2+1)-di\-men\-sion\-al (3D) effective theory of the parent vacuum with the compact dimension integrated out.  Compared to (\ref{N1}), this contains one fewer factor of the large ratio $\Hp/\Hinf$.  However, arguing along the same lines as above, it still seems possible that ${\cal N}_{\rm bub}\gtrsim {\cal O}(1)$.  

This paper explores the consequences of these ``anisotropic bubble collisions'' in terms of cosmological observables.  To simplify the analysis, we treat slow-roll inflation as a period of temporary vacuum-energy domination, which we turn off by hand after $N_e$ $e$-folds of expansion (i.e.~we do not treat the inflaton as a local degree of freedom).  This simplification would be insufficient for analyzing bubble collisions in the standard, entirely (3+1)-dimensional (4D) scenario, since in this case perturbations in the inflaton field are crucial to creating observable signatures of bubble collisions.  However, this is not the case in the present scenario, since gradients and time evolution of the tunneling modulus $\psi$ are observable features of the geometry.  Indeed, we find that anisotropic bubble collisions produce gravity waves, with the resulting CMB temperature and $E$- and $B$-mode polarization perturbations of each collision featuring a $\cos(2\phi)$ symmetry, where $\phi$ rotates the perturbation around the line-of-sight axis.  Whether or not these collisions also produce a significant scalar perturbation cannot be ascertained using our simple model, since our model ignores any 4D scalar degrees of freedom by construction.  However, because of the symmetries of the collision geometry we expect that any modification to the CMB temperature and $E$-mode polarization perturbations due to any scalar perturbation would be to simply add a radially symmetric perturbation (as in \cite{Czech:2010rg}) to the above $\cos(2\phi)$ rotationally symmetric profile.  Note that the $B$-mode polarization perturbation is independent of such considerations.

Direct searches for the effects of bubble collisions have so far focused on the temperature perturbations in the CMB, and they have constrained the size of these effects to be at most of order the amplitude of the inflationary perturbations \cite{Feeney:2010jj,Feeney:2010dd,Feeney:2012hj}.  It is therefore worth noting that the inflationary $B$-mode polarization perturbations (due to primordial gravity waves) have not yet been detected, and could be many orders of magnitude smaller than the inflationary scalar perturbations.  As such, the $B$-mode polarization effects of anisotropic bubble collisions could easily dominate over the inflationary ``background.''  Although the $E$-mode polarization from inflationary perturbations is gravitationally lensed into an apparent $B$-mode polarization perturbation by matter over-densities in the foreground, this signal can in principle be isolated and subtracted.
  
The remainder of this paper is summarized as follows.  In Section \ref{sec:modulus} we describe a simple model that implements metastable modulus decay.  An important feature of these transitions is that the tunneling instanton is homogeneous with respect to the ``decompactifying'' dimension $z$.  Thus, ignoring quantum perturbations, the $z$ dimension can be integrated out, yielding a 3D effective theory containing an additional scalar field, the modulus of the decompactifying dimension.  The situation is then analogous to the fully 4D scenario, containing a scalar inflaton, that is studied in most of the literature.  We close Section \ref{sec:modulus} with a brief description of the FRW cosmology in the 3D effective theory of a single bubble.  Then, in Section \ref{sec:EFT}, we obtain the general solution for a linear perturbation respecting the residual symmetry of an anisotropic bubble collision.  The actual perturbation from such a collision is obtained after introducing an appropriate ansatz for the modulus field profile in the wake of a collision, which we determine by analogy to the fully 4D scenario.  In Section \ref{sec:4D} we return to the full 4D picture of anisotropic bubble nucleation, computing the evolution of the resulting gravity waves.  The result is converted into CMB temperature and polarization observables in Section \ref{sec:signal}, and in Section \ref{sec:prospects} we discuss the prospects for observing these signals in terms of observational constraints and other considerations.  Concluding remarks are provided in Section \ref{sec:conclusion}.

\section{Metastable (de)compactification:  single bubble solution}
\label{sec:modulus}

\subsection{Toy model}
\label{ssec:toy}

Although the details of the microphysical model of metastable compactification are not important to our analysis, it will help to have a concrete picture in which to establish certain important results.  To this end we use the same (toy) model as \cite{BlancoPillado:2010uw} (and \cite{Salem:2010mi}).\footnote{Depending on the neutrino spectrum, the Standard Model might contain metastable 3D vacua with positive energy density, with the compact dimension stabilized by a combination of gauge-flux repulsion, Casimir-energy contraction, and vacuum-energy repulsion \cite{ArkaniHamed:2007gg}.  While potentially realistic as a low-energy effective theory, the model does not contain an inflaton and otherwise features too small of a vacuum energy density to serve our cosmological purpose.}  

In particular, we use the topological winding number of a non-canonical complex scalar field to stabilize the size of a compact dimension in the parent vacuum.  For simplicity we take the ``microphysical'' theory to be 4D (this is equivalent to ignoring any dimensions that are compact in our vacuum).  The action is
\beq
S = \int\! \sqrt{-g}\,d^4x \left[ \frac{1}{16\pi G}\left( R - 2\Lambda\right) 
-\frac{1}{2}K(X) - \frac{\lambda}{4}\left(|\varphi |^2-v^2\right)^2\right] ,
\label{4daction}
\eeq  
where $g$ is the determinant of the metric $g_{\mu\nu}$, $R$ is the Ricci scalar, and $\varphi$ is the complex scalar field for which we allow a non-canonical ``kinetic'' function specified by $K$, with
\beq
X\equiv\partial_\mu\varphi^*\partial^\mu\varphi \,.
\eeq  
The other terms in (\ref{4daction}) are constants.  Any additional degrees of freedom, for instance the inflaton and the fields of the Standard Model, are assumed to be unimportant in the parent vacuum and during the tunneling process and are absorbed into $\Lambda$ (and/or $g$ and $R$).  Note in particular that this means $\Lambda$ is not the late-time cosmological constant in our vacuum, but is instead dominated by the energy density of the inflaton.

The compactification of a (closed) spatial dimension $z$ is most conveniently studied using a metric ansatz with line element
\beq
ds^2 = e^{-\Psi}\,\overline{g}_{ab}\,dx^adx^b + L^2e^{\Psi}\,dz^2 \,,
\label{ansatz}
\eeq   
where $\Psi$ represents the modulus field.  The effective 3D metric $\overline{g}_{ab}$ and the modulus $\Psi$ are both taken to be independent of $z$.  Meanwhile, the coordinate $z$ is dimensionless and taken to obey periodic boundary conditions over the interval $-\pi < z < \pi$, so that it has the topology of a circle with physical circumference $2\pi L\,e^{\Psi/2}$, where $L$ is a constant with dimension length.  The topology of the other dimensions is unimportant, so long as they are sufficiently large.  Here and below Latin indices are understood to run over all but the $z$ dimension.

To proceed, consider for the moment that $\Psi$ is a constant, $\Psi=\Psi_{\rm p}$.  Then the equation of motion of $\varphi$ permits the winding solution
\beq
\varphi= \tilde{v}e^{inz} \,,
\label{phisol}
\eeq
where $n$ is an integer and $\tilde{v}$ is a constant, given by the solution to the algebraic equation
\beq
\tilde{v}^2 = v^2 - \frac{n^2}{\lambda L^2} K'_{\rm p}\,e^{-\Psi_{\rm p}} \,,
\label{veq}
\eeq
where $K'_{\rm p}\equiv dK/dX$, evaluated at $X(\Psi_{\rm p})=(n^2\tilde{v}^2/L^2)\,e^{-\Psi_{\rm p}}$.  The precise form of $\tilde{v}$ depends on the function $K$, since (\ref{veq}) contains a factor of $K'_{\rm p}$, which depends on $\tilde{v}$.  However, we can brush this complication aside by assuming
\beq
v^2 \gg \frac{n^2}{\lambda L^2}K'_{\rm p}\,e^{-\Psi_{\rm p}} \,, 
\label{etaapprox}
\eeq
so that to leading order we have $\varphi = ve^{inz}$.  Note that if (\ref{etaapprox}) is valid for $\Psi=\Psi_{\rm p}$, then it also holds for $\Psi>\Psi_{\rm p}$ (we assume that $K(X)$ contains no poles in $X$).  Using this it can be shown that (\ref{phisol}) is an approximate solution to the equations of motion even for time-dependent $\Psi$, so long as $\Psi>\Psi_{\rm p}$.

\begin{figure}[t!]
\begin{center}
\includegraphics[width=0.45\textwidth]{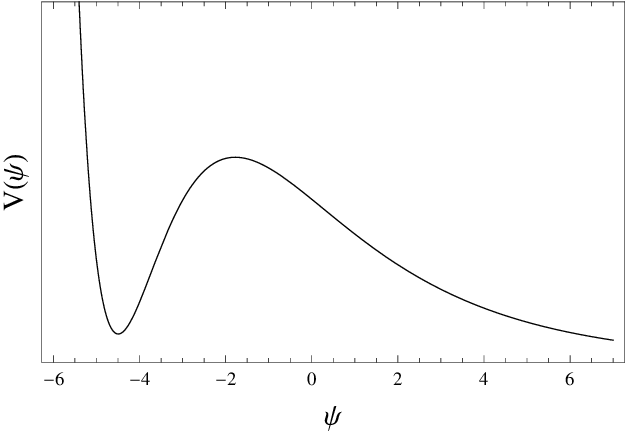} 
\caption{\label{fig:V}Example effective potential of the modulus field $\psi$.}
\end{center}
\end{figure}

The solution (\ref{phisol}) creates the desired metastable minimum for the modulus $\Psi$, given a suitable kinetic function $K$.  This is evident if we integrate out the $z$ dimension, generating (after integration by parts) the effective 3D action
\beq
\overline{S} = \int\! \sqrt{-\overline{g}}\,d^3x\left\{ 
\frac{1}{16\pi \overline{G}}\,\overline{R}
-\frac{1}{2}\partial_a\psi\partial^a\psi 
-\frac{\Lambda}{8\pi \overline{G}}\,e^{-\alpha\psi}
-\frac{1}{2}e^{-\alpha\psi} \overline{K}[X(\psi)]\right\}\,,
\label{3daction}
\eeq   
where we have defined $\overline{G}\equiv G/(2\pi L)$, $\overline{K}\equiv 2\pi L\, K$, and $\psi\equiv\Psi/\alpha$, with $\alpha\equiv\sqrt{16\pi\overline{G}}$.  Here and elsewhere overlines denote 3D quantities.  The rescaled modulus field $\psi$ now appears as a canonical scalar field, with effective potential
\beq
\overline{V}(\psi) = \frac{\Lambda}{8\pi \overline{G}}\,e^{-\alpha\psi}
+\frac{1}{2}e^{-\alpha\psi} \overline{K}[X(\psi)] \,.
\label{Veff}
\eeq
Choosing $\overline{K}(X) = 2\pi L(X+\kappa_2 X^2+\kappa_3 X^3)$, with appropriate constants $\kappa_2$, $\kappa_3$, $\Lambda$, etc., we obtain a modulus potential of the desired form, e.g.~that of Figure \ref{fig:V}.  (For additional details about this solution, see \cite{BlancoPillado:2010uw}.)

To be explicit:  the effective potential of Figure \ref{fig:V} has a positive-energy local minimum at some value of the modulus $\psi=\psi_{\rm p}$.  This solution corresponds to the parent vacuum.  Ac\-cord\-ing to the classical equations of motion of the 3D effective theory, if a sufficiently large volume is in a state $\psi(x)$ sufficiently close to $\psi_{\rm p}$, the local geometry will rapidly approach a patch of 3D de Sitter space with curvature radius $\Hp^{-1}$, where $\Hp^2 = 8\pi\overline{G}\,\overline{V}(\psi_{\rm p})$.  A bubble of our vacuum is created when in some localized region $\psi$ tunnels through the barrier, to some value $\psi=\psi_{\rm b}$.  Within the bubble, $\psi$ accelerates from rest and rolls down the potential, with $\psi\to\infty$ as time $x^0\to\infty$.  This corresponds to the decompactification of the $z$ dimension, since in the 4D picture the circumference of the $z$ dimension grows exponentially with $\psi$.  Note that the growth of the $z$ dimension does not limit the validity of the 3D effective theory, which integrates this dimension out.  This is because $\overline{g}_{ab}$ and $\psi$ are independent of $z$, which means the 3D effective theory is valid on all scales (up to perturbative quantum corrections).
 
Note that the decompactification decay described above is not the only decay channel of the parent-vacuum solution (\ref{phisol}).  In particular, it should be possible to nucleate a bubble in which the scalar winding number $n$ is reduced, in analogy to the flux-discharge decays in Einstein-Maxwell theory described in \cite{BlancoPillado:2009di,Carroll:2009dn}.  The bubble wall then corresponds to a charged 1-brane of scalar $\varphi$.  The effect of reducing the winding number is simply to lower the energy of the local minimum of the effective potential $\overline{V}(\psi)$, indicating that the $z$ dimension remains compactified within such a bubble (or, if the winding number is completely discharged, the decay creates a bubble of nothing \cite{BlancoPillado:2010df}).  Although we do not live in such a bubble, we might observe collisions between our bubble and these ones.  If the instantons describing these ``discharge'' decays are independent of $z$, as in the commonly used ``smeared brane'' assumption, then from the standpoint of our analysis these collisions are equivalent to collisions with decompactifying bubbles.  On the other hand, if the instantons depend on $z$, then collisions with discharge bubbles can in general give modified signatures.  For concreteness, we assume that we can choose the parameters of our toy model so that decompactification decays are exponentially more likely than discharge decays, so that we only observe the effects of collisions with decompactification bubbles.  Analogy to flux compactification in Einstein-Maxwell theory suggests that this is a plausible assumption \cite{Brown:2010mf}.

\subsection{Bubble geometry}
\label{ssec:sbs}

Within the context of the 3D effective theory, bubble formation can be understood in direct analogy to the work of \cite{Banks:1973ps,Coleman:1977py,Callan:1977pt,Coleman:1980aw,Brown:2007sd}.  In the ideal of a single bubble nucleation, the geometry surrounding the bubble wall is described by the line element
\beq
ds^2 = \oa^2(\eta)\Big[d\eta^2\!-\!d\xi^2+\cosh^2(\xi)\,d\phi^2\Big]\,.
\label{Lmetric}
\eeq
The SO(2,1) symmetry of this geometry is the reduced-dimension analogue of the SO(3,1) symmetry of the usual, fully 4D scenario of bubble nucleation.  The scale factor $\oa$, along with the tunneling modulus $\psi$, obeys the ``inverted potential'' equations of motion,
\beq
\frac{\dot{\oa}\phantom{\oa}\!\!\!^2}{\oa^2}-1
= 8\pi \overline{G} \left(\frac{1}{2}\dot{\psi}^2 
-\oa^2\,\overline{V}\right) \,, \qquad
\ddot{\psi} + \frac{\dot{\oa}}{\oa}\,\dot{\psi} 
= \oa^2\,\overline{V}' \,,\phantom{\bigg(\bigg)} \label{E}
\eeq
where the dot and prime denote differentiation with respect to $\eta$ and $\psi$ respectively.  These are different than the usual 4D equations only by a couple of numerical factors.  

\begin{figure}[t!]
\begin{center}
\includegraphics[width=0.33\textwidth]{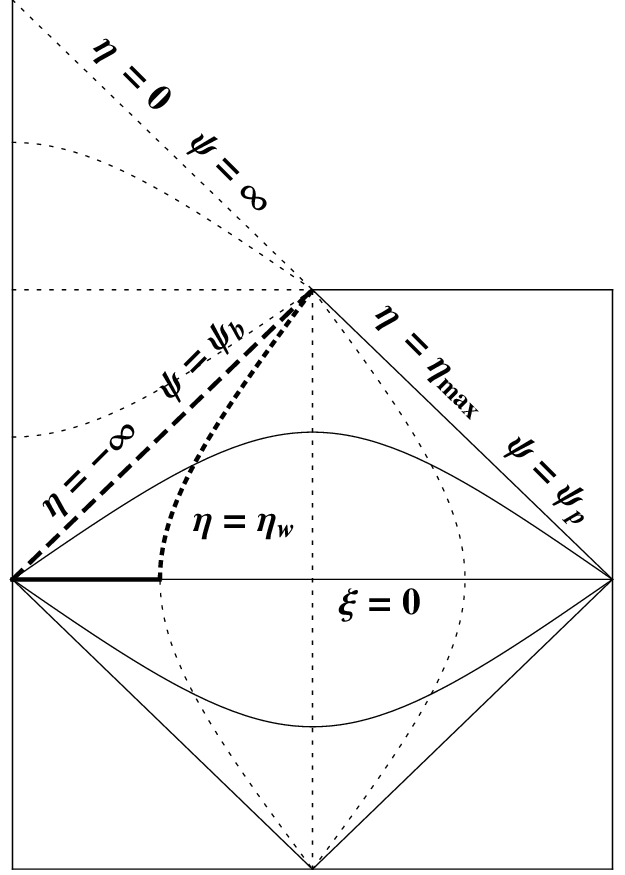} 
\caption{\label{fig:conformal}Conformal diagram of a single bubble nucleation in the 3D effective theory.  Each point on the diagram represents a circle, corresponding to rotating the diagram about its left edge.  The line element (2.9) describes the central diamond:  dotted curves are surfaces of constant $\eta$, solid curves are surfaces of constant $\xi$.  (The pre-bubble-nucleation geometry is matched onto the post-bubble-nucleation geometry at $\xi=0$.)  The initial bubble is a disk in 3D, represented by the thick horizontal line segment, and the trajectory of the bubble wall follows the thick dotted curve along $\eta=\eta_{\rm w}$ (surfaces of constant $\eta$ are surfaces of constant $\psi$).  As one moves from $\eta_{\rm w}$ to larger values of $\eta$, the geometry approaches de Sitter space with curvature radius $\Hp^{-1}$.  As one moves from $\eta_{\rm w}$ to smaller values of $\eta$, the geometry approaches de Sitter space with curvature radius $\Hb^{-1}$, where $\Hb^2 = 8\pi\overline{G}\,\overline{V}(\psi_{\rm b})$.  The limit $\eta\to-\infty$ corresponds to the future lightcone of the center of the initial bubble (thick dashed line); the region above this is described by the line element (2.11), which analytically continues the coordinates $\eta$ and $\xi$ (dotted curves are still surfaces of constant $\eta$).  Inside the bubble, $\psi$ evolves from $\psi_{\rm b}$ to $\infty$, which in the full 4D picture corresponds to the decompactification of the $z$ dimension.}
\end{center}
\end{figure}

The salient features of the geometry are summarized in Figure \ref{fig:conformal}.

The geometry within the future lightcone of the center of the initial bubble is obtained by solving the normal (Lorentzian) field equations of the 3D effective theory, using analytic continuation ($\oa\to i\oa$ and $\xi\to\xi-i\pi/2$) of the above solution to provide the in\-i\-tial conditions as $\eta\to-\infty$ in the bubble.  The analytic continuation takes $\eta$ to a timelike co\-or\-di\-nate and $\xi$ to a spacelike coordinate, and the resulting line element is
\beq
ds^2 = \oa^2(\eta)\Big[\!-\!d\eta^2+d\xi^2+\sinh^2(\xi)\,d\phi^2\Big]\,.
\label{bubblegeom0}
\eeq
The evolution of $\oa$ and $\psi$ depend on the matter content of the bubble.  For example, our toy model approximates slow-roll inflation as vacuum-energy domination, via the 4D cosmological constant $\Lambda$.  In the 3D effective theory this contributes a term to the effective potential $\overline{V}(\psi)$, namely the first term in (\ref{Veff}).  As $\psi$ increases this term quickly comes to dominate over the other terms, so that
\beq
\overline{V}(\psi) \to \frac{\Lambda}{8\pi \overline{G}}\,
e^{-\alpha\psi} \,, \label{Veff2}
\eeq
since we assume $\overline{K}(X)$ contains no poles in $X$ and $X\propto e^{-\alpha\psi}$.  In this limit we find
\beq
\overline{a}(\eta)= c_0\,\frac{\cosh(\eta)}{\sinh^2(\eta)} \,, \qquad
\psi(\eta) = \frac{1}{\sqrt{16\pi\overline{G}}}\ln\left[
\frac{\Lambda}{3}\,c_0^2\,\frac{\cosh^2(\eta)}{\sinh^2(\eta)}\right] ,
\eeq
where $c_0$ is a constant that is related to $\psi_{\rm b}$.  Our analysis focuses on times that are well after the onset of slow-roll inflation, and on scales that are small compared to the curvature radius.  Collectively, these correspond to $\Delta\xi\ll 1$ and $|\eta|\ll 1$, in which case the line element can be written  
\beq
ds^2 = \oa^2(\eta)\Big[\!-\!d\eta^2 + dx^2 + dy^2\Big] \,,
\label{gbubble}
\eeq
and the inflationary solution simplifies to
\beq
\oa(\eta) = \frac{c_0}{\eta^2}\,, \qquad
\psi(\eta) = \frac{1}{\sqrt{16\pi\overline{G}}}
\ln\!\left(\frac{\Lambda}{3}\frac{c_0^2}{\eta^2}\right) ,
\label{csolution}
\eeq
where in (\ref{gbubble}) we have switched to the Cartesian coordinates $\{x,y\}$ in lieu of $\{\xi,\phi\}$.  Note that the geometry of the 3D effective theory with scalar field $\psi$ and potential (\ref{Veff2}) is not equivalent to 3D de Sitter space.

\subsection{4D FRW evolution in the bubble}
\label{ssec:4DFRW}

To proceed beyond inflation in the bubble, it is easiest to return to the 4D picture in which the $z$ dimension is restored.  This is straightforward to do, since as we have stressed above the modulus $\psi$ is independent of $z$.  (This means, in particular, that the tunneling instanton is independent of $z$.  For example the initial bubble, which is a disk in the 3D effective theory, is a solid 2-torus wrapping the $z$ dimension in the full 4D picture.)  Referring to (\ref{ansatz}), the 4D line element within the bubble is
\bea
ds^2 &=& e^{-\alpha\psi(\eta)}\,\oa^2(\eta)
\Big[\!-\!d\eta^2+d\xi^2+\sinh^2(\xi)\,d\phi^2\Big] 
+ L^2e^{\alpha\psi(\eta)}\, dz^2 \,, \\
&=& a^2(\eta)\Big[\!-\!d\eta^2+d\xi^2
+\sinh^2(\xi)\,d\phi^2\Big] + b^2(\eta)\,dz^2 \,,
\label{bubblevac}
\eea
where in the second line we have defined $a^2=e^{-\alpha\psi}\,\oa^2$ and $b^2=L^2\,e^{\alpha\psi}$.  The scale factor $b$ can be viewed to take on the role of the modulus $\psi$; in particular the field equations give
\bea
\frac{\dot{a}^2}{a^2}+2\frac{\dot{a}\dot{b}}{ab}-1 &=& 8\pi G\,a^2\,V(b) \\
2\frac{\ddot{a}}{a}-\frac{\dot{a}^2}{a^2}-1 &=& 
8\pi G\,a^2\frac{d}{db}\big[bV(b)\big] \,,
\eea
where $V(b)=(1/2\pi L)\,e^{\alpha\psi(b)}\,\overline{V}[\psi(b)]$, with $\psi(b)=(2/\alpha)\ln(b/L)$ in accordance with above.  

In the 4D picture, the inflationary solution corresponds to
\beq
a(\eta) = -\frac{1}{\Hinf\sinh(\eta)} \,, \qquad
b(\eta) = -c_0L\Hinf\,\frac{\cosh(\eta)}{\sinh(\eta)} \,,
\eeq
where $\Hinf\equiv\sqrt{\Lambda/3}$.  This is locally equivalent to 4D de Sitter space.  In particular, well after the onset of slow-roll inflation and on scales that are small compared to the curvature radius, the line element can be written
\beq
ds^2 = a^2(\eta)\Big[\!-\!d\eta^2+dx^2+dy^2+d\tz^2\Big] \,,
\label{bubblevac2}
\eeq
where now 
\beq
a(\eta) = -\frac{1}{\Hinf\,\eta} \,, \qquad
\rho(\eta) = \frac{\Lambda}{8\pi G}=\frac{3\Hinf^2}{8\pi G}\,,
\eeq
and we have defined $\tz \equiv c_0L\Hinf^2\,z$.  We have noted the inflationary energy density for reference.  The conformal time after $N_e$ $e$-folds of inflation is
\beq
\eta_\star = -2e^{-N_e} \,.
\eeq

To describe the subsequent cosmology in the bubble, we forego the complexities of a microphysical description of the matter sector and instead simply solve the field equations for the appropriate perfect fluids.  Moreover, we assume instantaneous transitions between epochs during which only one cosmological fluid dominates, matching each epoch to the next by demanding that the scale factor be continuous and smooth at each transition.  For instance, we assume that inflation ends with instantaneous reheating at time $\eta=\eta_\star$, after which the vacuum energy $\Lambda$ is replaced by radiation and the scale factor and energy density are
\beq
a(\eta) = \frac{1}{\Hinf\,\eta_\star^2}\,\big(\eta-2\eta_\star\big) \,, 
\qquad
\rho(\eta) = \frac{3}{8\pi G\,\Hinf^2\,\eta_\star^4}\,
\frac{1}{a^4(\eta)} \,.
\label{sfr}
\eeq 
Whereas during inflation $\eta$ asymptotically approaches zero from below, during radiation domination $\eta$ smoothly passes through zero and becomes positive.  Radiation domination gives way to (non-relativistic) matter domination.  Again, our simplified model takes the transition to be instantaneous, at time $\eta=\eta_{\rm eq}$.  Since radiation domination (in bubbles like ours) lasts for many $e$-folds of cosmic expansion, $\eta_{\rm eq}\gg |\eta_\star|$, and we can approximate the subsequent scale factor and energy density solutions by
\beq
a(\eta) = \frac{1}{4\Hinf\,\eta_\star^2\,\eta_{\rm eq}}\,
\big(\eta+\eta_{\rm eq}\big)^2 \,, \qquad
\rho(\eta) = \frac{3}{8\pi G\,\Hinf\,\eta_\star^2\,\eta_{\rm eq}}\,
\frac{1}{a^3(\eta)}\,.
\label{sfm}
\eeq  
Since our observations occur relatively soon after the onset of the present dark-energy domination, at our level of analysis we can ignore the effects of the dark energy.  

The scale factor grows by a factor of over three thousand between $\eta_{\rm eq}$ and the present conformal time $\eta_{\rm obs}$; therefore we have $\eta_{\rm obs}\gg\eta_{\rm eq}\gg|\eta_\star|$.  At the same time, the size of $\eta_{\rm obs}$ can be related to the size of the present-day curvature parameter $\Omega_k$.  In particular, note that although the above analysis ignores the spatial curvature in our bubble, it maintains a normalization of the scale factor so that   
\beq
\Omega_k \equiv \frac{1}{H^2a^2} = \frac{a^2}{\dot{a}^2} \,,
\label{omega}
\eeq
where $H\equiv\dot{a}/a^2$ is the usual Hubble parameter (the unusual power on $a$ is because the dot denotes a derivative with respect to conformal time).  Combining (\ref{sfm}) with (\ref{omega}), we find
\beq
\eta_{\rm obs} \simeq 2\sqrt{\Omega_k} \,,
\label{etaobs}
\eeq
where $\Omega_k$ is understood to be evaluated at the present.  As mentioned in Section \ref{sec:introduction}, in the context of anisotropic bubble nucleation (and absent fine-tuning) we have $\Omega_k\lesssim 10^{-5}$.  Therefore we can safely take $\eta_{\rm obs}\ll 1$.

\section{Bubble collision in the 3D effective theory}
\label{sec:EFT}

We are unable to solve for the full geometry of an anisotropic bubble collision.  Nevertheless, the results of Section \ref{sec:modulus} indicate that in terms of the 3D effective theory with the $z$ dimension integrated out, an\-i\-so\-trop\-ic bubble nucleation looks just like ``isotropic'' bubble nucleation in the usual, fully 4D scenario, except for two differences.  In the usual, fully 4D scenario, the bub\-ble geometry has SO(3,1) symmetry, and there is a great deal of model-selection freedom when specifying the potential of the tunneling field, which for simplicity can also serve as the inflaton.  However, with an anisotropic bubble in the context of the 3D effective theory, the bubble geometry has only SO(2,1) symmetry, and while there is still a great deal of model-selection freedom to specify the potential of the tunneling modulus $\psi$ in the vicinity of the tunneling barrier, the shape of its potential for values of $\psi$ well inside the bubble is dictated by the dimensional reduction.  The first difference simply corresponds to the elimination of a coordinate from within a highly symmetric sub-manifold, and so we expect there to be a simple relationship between the two scenarios in this respect.  Meanwhile, the second difference only concerns the propagation of effects well into the bubble.

The effects of isotropic bubble collisions in the usual, fully 4D scenario are well-studied, both numerically and analytically (Section \ref{sec:introduction} contains a list of references).  Indeed, the recent analysis by Gobbetti and Kleban \cite{Gobbetti:2012yq} is particularly amenable to applying the analogy outlined above.  Gobbetti and Kleban solved for the propagation of a generic perturbation---consistent with the residual symmetry of the bubble collision---in the inflaton field, and matched the perturbation to the ``initial'' condition expected from a bubble collision.  The calculation is simplified by the fact that when computing potentially observable effects, one can focus on scales that are small compared to the spatial curvature radius in the bubble.  We can perform the analogous analysis in the 3D effective theory of Section \ref{sec:modulus}.  In the end, it is straightforward to translate the results to the 4D picture relevant to making cosmological predictions. 

As described in Section \ref{sec:modulus}, the nucleation of one bubble breaks the SO(3,1) symmetry of de Sitter space (in the 3D effective theory) down to SO(2,1).  Similarly, the nucleation of a second bubble, so that the two may collide, breaks the SO(2,1) symmetry down to SO(1,1).  This is easiest to visualize if we analytically continue to Euclidean space, where the 3D de Sitter space has O(4) symmetry.  According to the standard hypothesis \cite{Coleman:1980aw}, bubble nuc\-le\-a\-tion preserves as much symmetry as possible, consistent with the bubble being placed at some random spot on the 3-sphere.  The first bubble therefore breaks the O(4) symmetry down to O(3)---corresponding to the symmetry-preserving rotations of the 3-sphere about the ran\-domly placed center of the bubble---while the second bubble breaks the O(3) symmetry down to O(2)---corresponding to the symmetry-preserving rotations of the 3-sphere about the axis connecting the centers of two randomly placed bubbles.  Analytic continuation back to Lor\-ent\-zian signature (and choosing an orientation for the positive direction of time) gives SO(1,1) symmetry.  For simplicity we focus on the case of a single bubble collision.  Since we assume the effects of the collision are linear perturbations to the background geometry, the situation with many collisions can be understood via superposition.\footnote{\label{foot:iso}In fact, our assumptions require only that each perturbation is linear at the time that we establish ``initial'' conditions for the perturbation, i.e.~after the onset of slow-roll inflation in the bubble, as described later in the text.  In particular, the perturbation may be non-linear near the bubble wall.  If this is the case, and if multiple collisions overlap at the bubble wall, then the resulting late-time perturbation would presumably be modified relative to a simple linear superposition of perturbations as if each collision occurred in isolation.  Nevertheless, while the causal futures of several bubble collisions might feature some overlap in their intersections with the surface of last scattering, it is much less likely that the collisions overlap in their intersections with the bubble wall.  Indeed, this becomes probable only as the decay rate per unit Hubble volume of the parent vacuum approaches unity \cite{Salem:2011qz}.  This validates the superposition argument above.}

Before accounting for any bubble collisions, the local line element in the bubble is given by (\ref{gbubble}).  As we have remarked, we focus on scales that are small compared to the curvature radius, and we focus on regions in the bubble for which the effects of any collisions are only small perturbations to the background.  We are free to orient the $\{x,y\}$ coordinates of the line element (\ref{gbubble}) so that the coordinate $y$ runs along the ``1-dimensional hyperboloid'' (1-sphere in Euclidean space) whose symmetry is preserved by the collision.  The most general perturbation consistent with this symmetry depends only on $\eta$ and $x$.  Accordingly, we write
\beq
\psi(\eta,x) = \frac{1}{\sqrt{16\pi\overline{G}}}\,
\ln\!\left\{\frac{\Lambda}{3}\frac{c_0^2}{\eta^2}
\Big[1+\delta\psi(\eta,x)\Big]\right\} ,
\eeq
and we assume $\delta\psi\ll 1$.  

Any modulus perturbation $\delta\psi$ will back-react on the geometry to produce a metric perturbation $\delta\overline{g}_{ab}$.  Since in the 4D picture $\delta\psi$ and $\delta\overline{g}_{ab}$ correspond to metric perturbations of the same order, we cannot self-consistently ignore the back-reaction of $\delta\psi$ on the 3D geometry.  Therefore, we must develop cosmological perturbation theory in 3D.  The most general linearly perturbed 3D FRW line element can be written
\bea
ds^2 &=& \oa^2(\eta)\Big\{ -\!\big[1+2\overline{\phi}(\eta,x,y)\big] 
d\eta^2 +2\big[\partial_i \overline{B}(\eta,x,y)
-\overline{S}_i(\eta,x,y)\big] d\eta\,dx^i \Big. \nn\\*
& & + \Big.\,
\big[1-2\overline{\zeta}(\eta,x,y)\,\delta_{ij} 
+ 2\,\partial_i\partial_j \overline{E}(\eta,x,y) 
+\partial_i\overline{F}_j(\eta,x,y)
+\partial_j\overline{F}_i(\eta,x,y)\big] 
dx^i\,dx^j \Big\} \,, \quad\,\,\,
\label{metric2}
\eea
where the indices run over the coordinates $\{x,y\}$.  This has been constructed in analogy to the usual linearly perturbed 4D FRW metric, as described in e.g.~\cite{Mukhanov:1990me}.  Here $\overline{\phi}$, $\overline{B}$, $\overline{\zeta}$, and $\overline{E}$ are ``scalar'' perturbations and $\overline{S}_i$ and $\overline{F}_i$ are the components of 3D ``vector'' perturbations.  The vector perturbations are understood to be subject to the constraints $\partial_i\overline{S}^i=\partial_i\overline{F}^i=0$.  It is straight\-forward to compare the number of perturbations to the number of degrees of free\-dom in a 3D symmetric tensor and verify that we have captured them all.  Although the 3D effective theory does not contain an analog of tensor perturbations, this of course does not imply that it does not excite tensor perturbations in the 4D picture.
   
In 4D, vector perturbations are not sourced by scalar-field fluctuations, and the same is true here.  In particular, in the 3D effective theory the modulus $\psi$ is a scalar field, and so its perturbations do not source the 3D vector perturbations.  Therefore we set $\overline{S}_i=\overline{F}_i=0$.  

Moreover, among $\overline{\phi}$, $\overline{B}$, $\overline{\zeta}$, $\overline{E}$, and $\delta\psi$, there is only one scalar degree of freedom; the rest are related by diffeomorphism (gauge) invariance and constraints coming from the equations of motion.  In particular, under the generic ``scalar'' infinitesimal coordinate transformation
\beq 
\eta\to\eta+\alpha(\eta,x,y) \,, \quad x^i\to x^i + \partial^i\beta(\eta,x,y) \,,
\label{redefinitions}
\eeq
the scalar metric perturbations transform according to 
\beq
\overline{\phi}\to\overline{\phi}-\dot{\alpha}-\overline{\cH}\alpha\,, \quad\,
\overline{B}\to\overline{B}+\alpha-\dot{\beta}\,, \quad\,
\overline{\zeta}\to\overline{\zeta}+\overline{\cH}\alpha\,, \quad\,
\overline{E}\to\overline{E}-\beta \,,
\label{newgauges}
\eeq
where $\overline{\cH}\equiv\dot{\oa}/\oa$.  We can thus fix the gauge by setting $\overline{B}=\overline{E}=0$, which is always possible by setting $\alpha=-\overline{B}+\dot{\overline{E}}$ and $\beta=\overline{E}$.  This is the analog of longitudinal gauge.  In 4D, the field equations in this gauge would set the 4D analog of $\overline{\phi}$ equal to the 4D analog of $\overline{\zeta}$.  In 3D, the analogous analysis finds that $\overline{\phi}=0$.  Therefore, the most general (gauge-fixed) metric (with only a scalar field source) can be written
\beq
ds^2 = \oa^2(\eta)\Big\{\!-\!d\eta^2 
+ \big[1-2\overline{\zeta}(\eta,x,y)\big]\big(dx^2+dy^2\big) \Big\} \,.
\eeq    

We can now solve for the most general set of perturbations, consistent with the SO(1,1) symmetry of an anisotropic bubble collision.  The $(\eta,\eta)$ and $(\eta,x)$ components of the Einstein field equations respectively give
\bea
3\delta\psi(\eta,x)+\eta\,\delta\dot{\psi}(\eta,x)+4\eta\,\dot{\zeta}(\eta,x)
+\eta^2\partial_x^2\zeta(\eta,x) &=& 0 \phantom{\big[\big]}\\
\partial_x\delta\psi(\eta,x)+\eta\,\partial_x\dot{\zeta}(\eta,x) &=& 0 \,, \phantom{\big[\big]}
\eea
which are redundant with the other components of the field equations and with the equation of motion of $\delta\psi$.  Here we have input the background solution (\ref{csolution}), and we have used the SO(1,1) symmetry to eliminate the $y$-dependence of $\delta\psi$ and $\zeta$.  The solution can be written
\bea
\delta\psi(\eta,x) &=& \eta\,\big[ f'(x+\eta)-g'(x-\eta) \big] 
\label{gen1}\\
\overline{\zeta}(\eta,x) &=& -f(x+\eta)-g(x-\eta)  \label{gen2}\,,
\eea
where $f$ and $g$ denote arbitrary functions of a single argument, and the primes denote derivatives with respect to that argument.  To obtain the explicit functional forms of $f$ and $g$, we need to input the appropriate ``initial'' conditions established by a bubble collision.  For this we continue to follow the analysis of Gobbetti and Kleban, guided by the aforementioned similarities between anisotropic bubble collisions as they are expressed in the 3D effective theory and isotropic bubble col\-li\-sions in the usual, fully 4D scenario.    

We establish the initial conditions at some time $\eta=\eta_0$, taken to be after the onset of slow-roll inflation in the bubble (so that the background solution (\ref{csolution}) is valid).  For con\-crete\-ness, we take the boundary of the causal future of the collision to be incoming from pos\-i\-tive $x$, and we set the origin of the $x$ axis so that this boundary is located at $x=-\eta_0$ when $\eta=\eta_0$.  This makes the trajectory of the incoming boundary a simple function of $\eta$,  
\beq
\xc(\eta) = -\eta \,. 
\eeq
By causality, the perturbation in $\psi$ at $\eta=\eta_0$ is proportional to a step function centered at $\xc(\eta_0)=-\eta_0$, i.e.~$\delta\psi(\eta_0,x)\propto\Theta(x+\eta_0)$, where the proportionality factor can depend on $x$.  Since we are ultimately interested in small comoving scales, we can expand in $x$:
\bea
f(x+\eta_0) &=& \sum_{n=1}^{\infty} c_n\, (x+\eta_0)^n\,\Theta(x+\eta_0) \\
g(x-\eta_0) &=& \sum_{n=1}^{\infty} d_n\, (x+\eta_0)^n\,\Theta(x+\eta_0) \,.
\eea
Note that the sums starts at $n=1$, as opposed to $n=0$.  This is because the derivatives $f'$ and $g'$ appear in $\delta\psi$, and for these the $n=0$ terms would give divergent contributions to the gradient energy density in the delta function resulting from differentiating the step function.  Given the functional forms of $f$ and $g$, it is now trivial to insert the time dependence:
\bea
f(x+\eta) &=& \sum_{n=1}^{\infty} c_n\,(x+\eta)^n\,\Theta(x+\eta) \label{expansion1}\\
g(x-\eta) &=& \sum_{n=1}^{\infty} d_n\,(x-\eta+2\eta_0)^n\,\Theta(x-\eta+2\eta_0) \,. \label{expansion2}
\eea

\begin{figure}[t!]
\begin{center}
\includegraphics[width=0.6\textwidth]{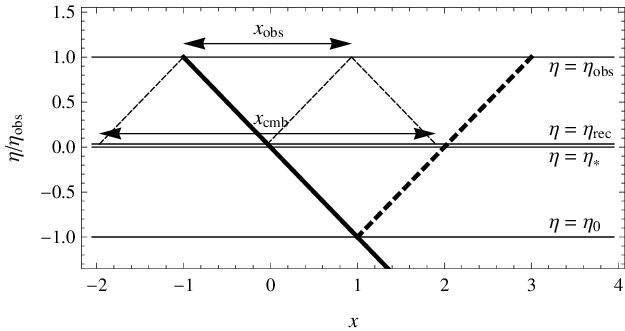} 
\caption{\label{fig:coll}Diagram of the spacetime surrounding the incoming boundary of the causal future of a collision, represented by the thick solid line corresponding to $\xc=-\eta$.  For simplicity we have chosen $\eta_0=-\eta_{\rm obs}$, and all quantities are expressed in units of $\eta_{\rm obs}$ (the diagram is not perfectly to scale).  The arrows under $x_{\rm obs}$ indicate the allowed values of $x_{\rm obs}$, consistent with the observer seeing the incoming wave front on her CMB.  The arrows under $x_{\rm cmb}$ indicate the range of points probed by the CMB, depending on $x_{\rm obs}$.  The thick dotted line is represents $x-\eta+2\eta_0=0$.}
\end{center}
\end{figure}

There are an infinite number of free parameters in (\ref{expansion1}) and (\ref{expansion2}), however only one of them, $c_1$, is important to our analysis.  This is because we are primarily interested in collisions for which the boundary of the causal future of the collision intersects the surface of last scattering.  In particular, for an observer located at $x=x_{\rm obs}$, any $x$ coordinate $x_{\rm cmb}$ on the surface of last scattering must satisfy
\beq
x_{\rm obs}-\eta_{\rm obs}+\eta_{\rm rec} \leq x_{\rm cmb} \leq x_{\rm obs}+\eta_{\rm obs}-\eta_{\rm rec} \,,
\label{xcmb}
\eeq
where $\eta_{\rm rec}$ is the time of recombination.  (Figure \ref{fig:coll} displays a diagram of the spacetime.)  At this time the boundary of the causal future of the collision is at $\xc = -\eta_{\rm rec}$; inserting this into (\ref{xcmb}) and rearranging some terms, we obtain 
\beq
-\eta_{\rm obs}\leq x_{\rm obs}\leq \eta_{\rm obs}-2\eta_{\rm rec} \,.  
\label{inequality}
\eeq
Since $\eta_{\rm rec}\ll\eta_{\rm obs}\ll 1$, we conclude that $|x_{\rm obs}|\ll 1$ and likewise that $|x_{\rm cmb}|\ll1$.  Therefore, we have $x+\eta\ll 1$ for all $x$ and $\eta$ of interest, and so we only need to keep the first term in (\ref{expansion1}).  With respect to (\ref{expansion2}), each term is proportional to a step function with argument $x-\eta+2\eta_0$.  With an appropriate choice of $\eta_0$, we can make this argument negative for all $x$ and $\eta$ of interest.  For instance, at the time of recombination the largest value this argument takes is $2\eta_{\rm obs}-4\eta_{\rm rec}+2\eta_0$, which comes from combining the right-hand-sides of (\ref{xcmb}) and (\ref{inequality}).  This is negative for $\eta_0<\eta_{\rm rec}-\eta_{\rm obs}$.  Referring to Figure \ref{fig:coll}, we see that if these step functions are zero at recombination, they are zero for all $x$ and $\eta$ of interest.

The above argument focuses on collisions for which the boundary of the causal future of the collision intersects the surface of last scattering.  Nevertheless, it is easy to see that it applies to many other collisions as well.  In particular, the first part only requires that $|x_{\rm obs}|\ll 1$.  The second part requires that $2\eta_0< x_{\rm obs}+\eta_{\rm obs}-\eta_{\rm rec}$, but the only constraint on $\eta_0$ is that it be set during slow-roll inflation.  The onset of inflation in the bubble is not abrupt, but note for example that the first slow-roll parameter is 1/10 when $\eta=-1/3$.  Thus, with good approximation we can set $\eta_0=-1/3$, in which case the second argument is also satisfied by $|x_{\rm obs}|\ll 1$.  Thus, we only need to keep track of the coefficient $c_1$ among (\ref{expansion1}) and (\ref{expansion2}) if only the boundary of the causal future of the collision is not a significant fraction of a curvature radius beyond the surface of last scattering.

Applying these results to (\ref{gen1}) and (\ref{gen2}), we obtain
\bea
\delta\psi(\eta,x) &=& c_1\eta\,\Theta(x+\eta) \label{sol1}\\
\overline{\zeta}(\eta,x) &=& -c_1(x+\eta)\,\Theta(x+\eta)  
\label{sol2}\,,
\eea
where we have dropped a term of the form $z\,\delta(z)$, which can formally be identified with zero.  The factor $c_1$ appears in our analysis as a free parameter, but in a more complete analysis it would relate to certain circumstances of the bubble collision, for instance the tensions in the bubble walls and/or the relative spacetime points at which the two colliding bubbles nucleated.  Lacking a more complete analysis, we simply require that $c_1$ satisfy $-|c_1|\eta_0\ll 1$ so as to justify the perturbative analysis of $\delta\psi$, with the smallness of $\overline{\zeta}$ then assured since $x+\eta$ is small (for all regions of interest).

\section{Bubble collision in the 4D picture}
\label{sec:4D}

The bubble-collision solution of Section \ref{sec:EFT} applies during slow-roll inflation in the bubble.  To proceed beyond inflation in the bubble, it is easiest to return to the full 4D picture with the $z$ dimension restored.  This is done using (\ref{ansatz}), exactly as was done in Section \ref{ssec:4DFRW}, but now including the (perturbative) effects of a bubble collision.  During slow-roll inflation in the bubble, this gives
\beq
ds^2 = \frac{1}{\Hinf^2\,\eta^2}\Big[\!-\!(1-\delta\psi)\,d\eta^2 +
(1-2\overline{\zeta}-\delta\psi)\big(dx^2+dy^2\big) 
+ (1+\delta\psi)\,d\tz^2 \Big] \,, \label{solution1}
\eeq
where $\delta\psi$ and $\overline{\zeta}$ are given by (\ref{sol1}) and (\ref{sol2}), and as before $\tz \equiv c_0L\Hinf^2\,z$.  We now express the metric perturbations in terms of 4D scalar, vector, and tensor perturbations.

The most general linearly perturbed 4D FRW line element can be written
\bea
ds^2 &=& a^2(\eta) \Big\{ -\!\big[1+2\phi(\eta,\bfx)\big] d\eta^2 
+2\big[\partial_i B(\eta,\bfx)- S_i(\eta,\bfx)\big] d\eta\,dx^i 
+ \big[1-2\zeta(\eta,\bfx)\,\delta_{ij} \quad \big. \Big.\nn\\* 
& & + \Big. \big. \,
2\,\partial_i\partial_j E(\eta,\bfx) +\partial_iF_j(\eta,\bfx)
+\partial_jF_i(\eta,\bfx)+h_{ij}(\eta,\bfx)\big] dx^i\,dx^j \Big\} \,,
\label{metric3}
\eea
where $\bfx=\{x,y,\tz\}$ and the indices now run over all three of these coordinates \cite{Mukhanov:1990me}.  Here $\phi$, $B$, $\zeta$, and $E$ are scalar perturbations, $S_i$ and $F_i$ are the components of 4D vector perturbations, and $h_{ij}$ are the components of tensor perturbations.  The vector perturbations are understood to be subject to the constraints $\partial_iS^i=\partial_iF^i=0$, while the tensor perturbations are subject to the constraints $h^i_{\phantom{i}i}=0$ and $\partial_ih^i_{\phantom{i}j}=0$.  

It is easily checked that the above bubble-collision solution is obtained by setting
\bea
\phi &=& - \frac{c_1}{2}\,\eta\,\Theta(x+\eta) \label{phi2}\\
\zeta &=& - \frac{c_1}{2}\,(x+\eta)\,\Theta(x+\eta) \\
E &=& \frac{c_1}{12}\Big[(x+\eta)^3-3\eta\,(x+\eta)^2\Big]\, \Theta(x+\eta) \label{E2}\\
h_{yy} &=& -h_{zz}\equiv h_+ = c_1\,x\,\Theta(x+\eta) \label{hinf}\,,\phantom{\frac{c}{4}}
\eea
with all of other metric perturbations set to zero.  Since we have treated slow-roll inflation as vacuum-energy domination and ignored inflaton perturbations, we should find that this is a solution to the field equations for pure cosmological constant, and it can be checked that this is indeed the case.  It is convenient to transform to longitudinal gauge, i.e.~to transform so as to set $B=E=0$.  The 4D scalar metric perturbations transform in exact analogy to the 3D transformations described in Section \ref{sec:EFT}, i.e.~they obey (\ref{newgauges}) under the transformations (\ref{redefinitions}), after removing the overlines and interpreting the index $i$ to run over $\{x,y,\tz\}$.  Choosing $\alpha$ and $\beta$ in those transformations so as to set $B=E=0$, $\phi$ and $\zeta$ also become zero.  Thus the scalar perturbations (\ref{phi2})--(\ref{E2}) above are simply gauge artifacts, and the anisotropic bubble collision produces a pure gravity wave in the full 4D picture.  

In hindsight, this result is not surprising.  We have treated slow-roll inflation as vacuum-energy domination---in particular we have not included an inflaton field degree of freedom in our analysis---so any scalar perturbations must be pure gauge.  So, while we find that it might be possible for an anisotropic bubble collision to produce no 4D scalar metric perturbations, we cannot assert that these perturbations are not produced.  Exploring this issue further requires a more detailed microphysical model describing the coupling between modulus and inflaton degrees of freedom, as well as a careful analysis of the microphysical dynamics of the bubble collision, and is beyond the scope of this paper.\footnote{Of course, in our linearized analysis, any scalar inflaton and metric perturbations would be decoupled from the tensor perturbations generated by an anisotropic bubble collision.  However, while we assume a linearized analysis is appropriate for times $\eta\geq\eta_0$, we do not presume a linear analysis is valid at all times.  It is therefore possible that the initial effects of a collision are non-linear, correlating tensor and scalar perturbation that have become linear by $\eta=\eta_0$.}  We therefore proceed by simply focusing on the tensor perturbation discovered above.

As described in Section \ref{ssec:4DFRW}, we assume that inflation gives way to instantaneous reheating followed by radiation domination.  We solve for the evolution of the metric perturbations (in the background of Section \ref{ssec:4DFRW}) by demanding that the perturbations be continuous and smooth at the transition.  The only non-zero perturbation is of course the tensor perturbation $h_{ij}={\rm diag}\{0,h_+,-h_+\}$.  During radiation domination, a generic perturbation $h_+$ (consistent with the symmetry of an anisotropic bubble collision) can be written
\beq
h_+(\eta,x)=\frac{1}{\eta-2\eta_\star}\,\big[ f(x+\eta) + g(x-\eta)\big] \,,
\eeq
where as before $f$ and $g$ are arbitrary functions of a single argument.  Matching this function and its first derivative onto the inflationary solution (\ref{hinf}) at $\eta=\eta_\star$, we find
\bea
h_+(\eta,x) &=& \frac{c_1}{\eta-2\eta_\star}\bigg\{
\frac{1}{4}\Big[(x+\eta)^2\,\Theta(x+\eta)-(x-\eta+2\eta_\star)^2\,\Theta(x-\eta+2\eta_\star) \Big] \bigg. \quad \nn\\*
& & \bigg. -\eta_\star(x+\eta-\eta_\star)\,\Theta(x+\eta) \bigg\} \\ 
&\to& \frac{c_1}{4\eta}\Big[(x+\eta)^2\,\Theta(x+\eta)-(x-\eta)^2\,\Theta(x-\eta)\Big]\,,
\eea
where the last line takes the limit of late times during radiation domination, using $\eta\gg|\eta_\star|$.

We assume radiation domination instantly gives way to (non-relativistic) matter domination, at $\eta=\eta_{\rm eq}$.  During matter domination, a generic perturbation $h_+$ can be written
\beq
h_+(\eta,x)=\frac{1}{(\eta+\eta_{\rm eq})^3}\,\Big\{ f(x+\eta)+g(x-\eta) 
-(\eta+\eta_{\rm eq}) \big[ f'(x+\eta) - g'(x-\eta)\big]\Big\} \,,
\label{hsol}
\eeq
where the primes denote derivatives with respect to the arbitrary functions $f$ and $g$.  Demanding that $h_+$ is continuously differentiable at $\eta=\eta_{\rm eq}$, we find
\bea
f(w) &=& -\frac{c_1}{480\eta_{\rm eq}}\Big[
w^3\big(w^2+20\eta_{\rm eq}w+160\eta_{\rm eq}^2\big)\,\Theta(w) 
-\big(w-2\eta_{\rm eq}\big)^5\,\Theta(w-2\eta_{\rm eq})\Big] 
\label{fsol} \\
g(w) &=& -\frac{c_1}{480\eta_{\rm eq}}\Big[
w^3\big(w^2-20\eta_{\rm eq}w+160\eta_{\rm eq}^2\big)\,\Theta(w) 
-\big(w+2\eta_{\rm eq}\big)^5\,\Theta(w+2\eta_{\rm eq})\Big] \,.
\quad\,\, \label{gsol}
\eea 
The result is complicated, but straightforward to analyze; see Section \ref{sec:signal}.  For the moment we simply note that a dramatic simplification occurs when $x\geq\eta\geq 2\eta_{\rm eq}$, in which case 
\beq 
h_+(\eta,x)= c_1 x \qquad \mbox{(for $x\geq\eta\geq\eta_{\rm eq}$)} \,.
\label{hnot}
\eeq

\section{CMB signatures}
\label{sec:signal}

We now express the effects of an anisotropic bubble collision in terms of CMB observables.  We stress that our simple cosmological model, which treats slow-roll inflation in our bubble as if it were simply a temporary period of vacuum-energy domination, does not include scalar metric perturbations by construction.  If such perturbations are generated in a more realistic model of these bubble collisions, then their observational consequences should be included alongside the effects we describe below.

\subsection{Temperature perturbation}
\label{ssec:temp}

Consider a beam of CMB photons traveling in the direction $\hat{n}$.  A tensor metric perturbation $h_{ij}$ induces a temperature fluctuation $\Theta\equiv\delta T/T_{\rm avg}$ according to (see for example \cite{Mukhanov:2005sc})
\beq
\Theta(\hat{n}) = -\frac{1}{2}\int_{\lambda_{\rm emit}}^{\lambda_{\rm obs}} d\lambda\,
\frac{\partial h_{ij}}{\partial \lambda} \hat{n}^i \hat{n}^j\,,
\label{tempani}
\eeq
where the integration follows a photon affine parameter $\lambda$ from the point of emission to the point of observation, and $T_{\rm avg}$ is the average temperature of the CMB.  In general, the metric perturbation also deflects the beam of photons with respect to the direction $\hat{n}$.  However, the effects on CMB observables from this deflection are suppressed by an additional factor of the metric perturbation.  We therefore ignore the deflection of $\hat{n}$ and focus on the red\-shift/blue\-shift of photon energies computed in (\ref{tempani}).

To proceed, it is convenient to switch to a spherical-polar coordinate system, with $\theta=0$ corresponding to the positive $x$ axis of our previous Cartesian system.  Our metric perturbation is of the form $h_{yy} = -h_{zz} = h_+(\eta,x)$, with all of the other components of $h_{ij}$ equal to zero.  Therefore, the temperature fluctuation seen by an observer at $\{\eta_{\rm obs},x_{\rm obs}\}$, looking in the direction $\hat{n} = \{\cos(\theta),\sin(\theta)\cos(\phi),\sin(\theta)\sin(\phi)\}$, is
\beq
\Theta(\theta,\phi) = -\frac{1}{2}\,
\Delta h_+(\theta)\left(\hat{n}_y^2 - \hat{n}_z^2\right)
= -\frac{1}{2}\,\Delta h_+(\theta)\,\sin^2(\theta)\cos(2\phi) \,,
\label{Thetadef} 
\eeq
where $\Delta h_+$ denotes the difference in $h_+$ between the points of observation and of emission,
\bea
\Delta h_+(\theta) = h_+(\eta_{\rm obs},x_{\rm obs})
-h_+\!\big[\eta_{\rm rec},x_{\rm obs}+(\eta_{\rm obs}-\eta_{\rm rec})\cos(\theta)\big]\,,
\label{eqn:Temp}
\eea
and we have set the time of emission to be the time of recombination (recall that we focus on perturbations that are incoming from the positive $x$ direction).  Note that all observers see a temperature fluctuation of the form $f(\theta)\cos(2\phi)$, a consequence of the $y$ and $z$ translation invariance of the metric perturbation.  In terms of the spherical-harmonic moments
\beq
\Theta_{\ell m} \equiv \int d\Omega\,\Theta(\theta,\phi)\,Y^*_{\ell m}(\theta,\phi) \,,
\eeq
only the $m=\pm 2$ moments are non-zero, and these are real and equal to each other.

\begin{figure}[t!]
\begin{center}
\begin{tabular}{ccc}
\quad\,\,\,\,\rule[0.5ex]{6pt}{0.55pt} {\scriptsize $\displaystyle 
\Theta(\theta,\phi)/[c_1\eta_{\rm obs}\cos(2\phi)]$} &
\begin{tabular}{c} 
\quad\,\,\rule[0.5ex]{6pt}{0.55pt} {\scriptsize $\displaystyle 
10^4 Q(\theta,\phi)/[c_1\eta_{\rm obs}\cos(2\phi)]$} \\
\vspace{-15pt}\\
\quad\,\,\rule[0.5ex]{0.75pt}{0.55pt}\hspace{1pt}\rule[0.5ex]{0.75pt}{0.55pt}\hspace{1pt}\rule[0.5ex]{0.75pt}{0.55pt}\hspace{1pt}\rule[0.5ex]{0.75pt}{0.55pt} {\scriptsize $\displaystyle 
10^4 U(\theta,\phi)/[c_1\eta_{\rm obs}\sin(2\phi)]$} 
\end{tabular} & 
\begin{tabular}{c} 
\quad{\scriptsize $\displaystyle \mbox{\large\textbullet}\,\,
10^8[\ell(\ell+1)/2\pi]C^E_\ell/(c_1\eta_{\rm obs})$} \\
\vspace{-15pt}\\ 
\quad{\scriptsize $\displaystyle \blacktriangle\,\, 
10^8[\ell(\ell+1)/2\pi]C^B_\ell/(c_1\eta_{\rm obs})$} 
\end{tabular} \\
\includegraphics[width=0.29\textwidth]{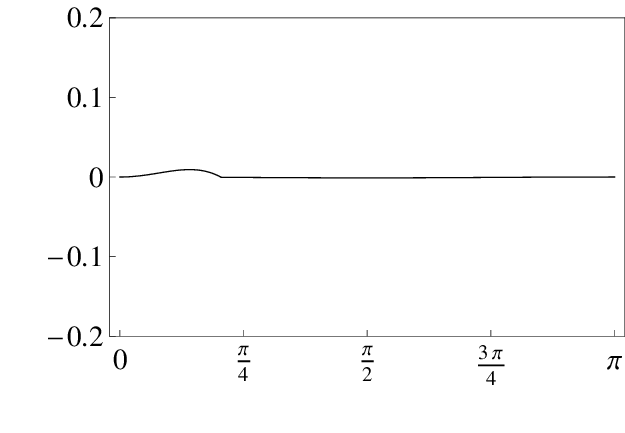} & 
\includegraphics[width=0.29\textwidth]{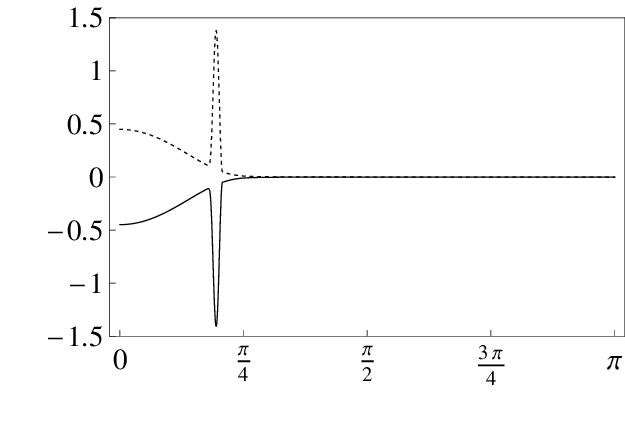} & 
\includegraphics[width=0.29\textwidth]{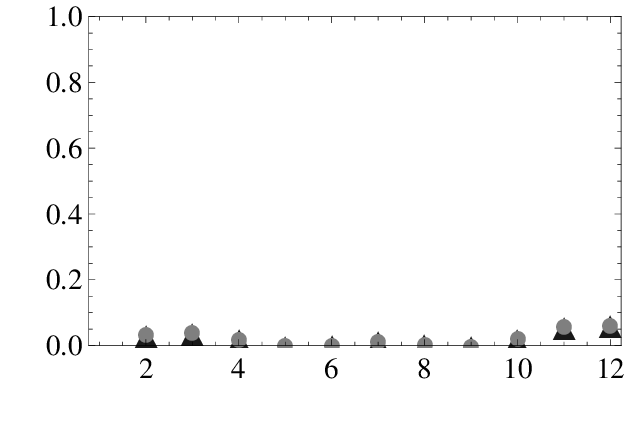} \\
\includegraphics[width=0.29\textwidth]{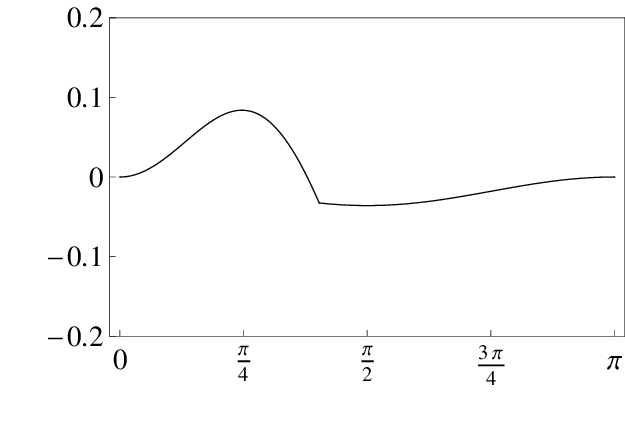} & 
\includegraphics[width=0.29\textwidth]{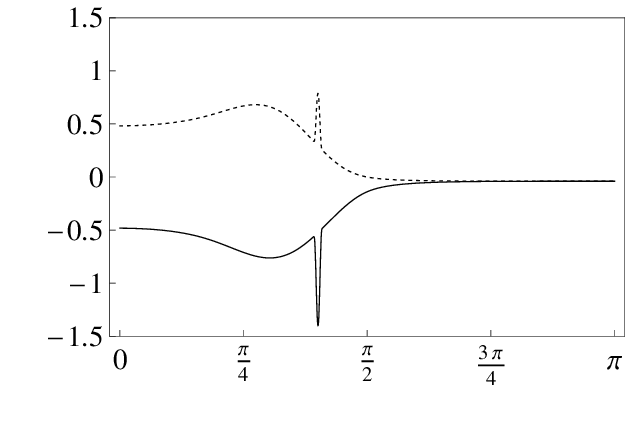} & 
\includegraphics[width=0.29\textwidth]{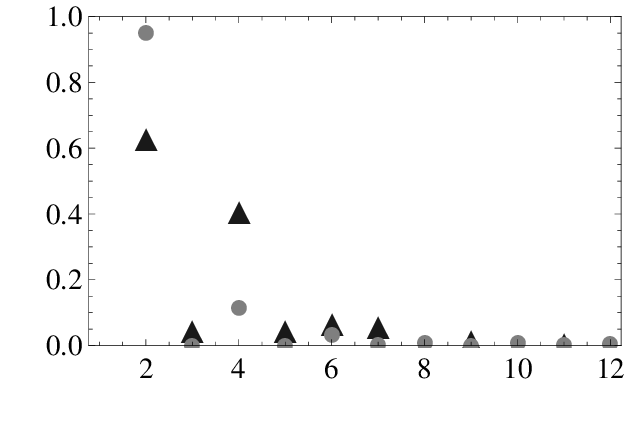} \\
\includegraphics[width=0.29\textwidth]{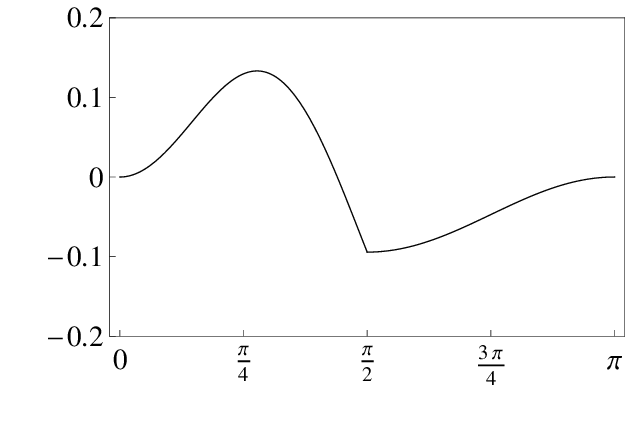} & 
\includegraphics[width=0.29\textwidth]{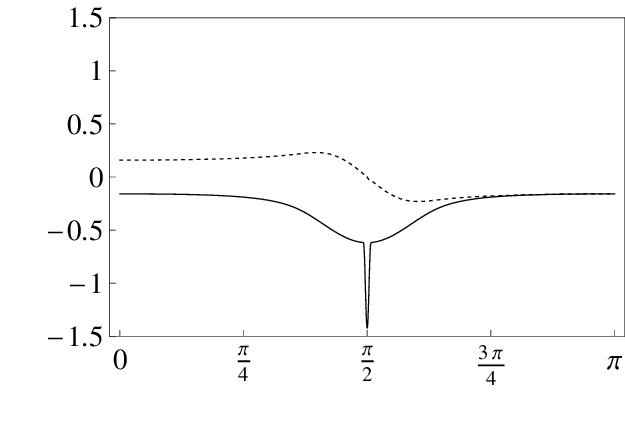} & 
\includegraphics[width=0.29\textwidth]{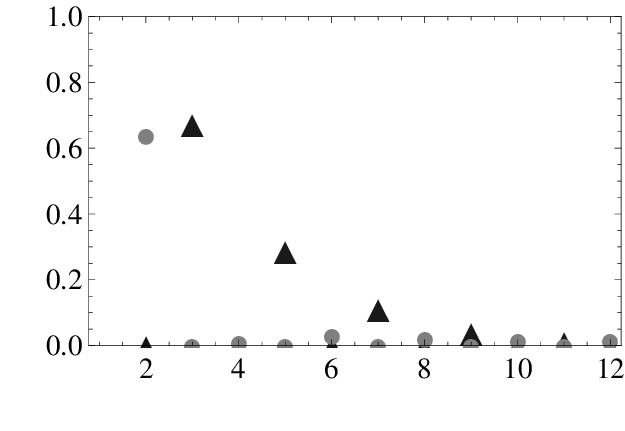} \\
\includegraphics[width=0.29\textwidth]{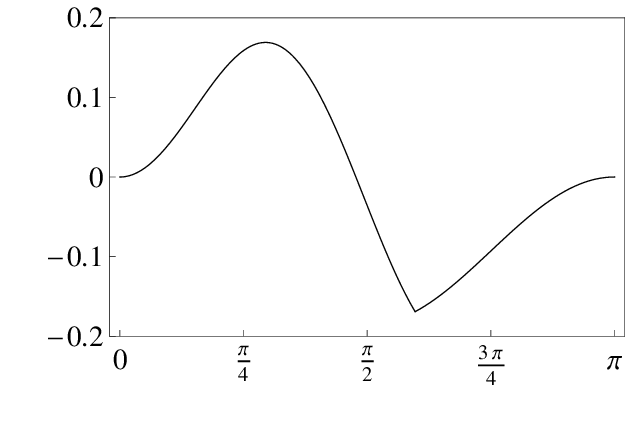} & 
\includegraphics[width=0.29\textwidth]{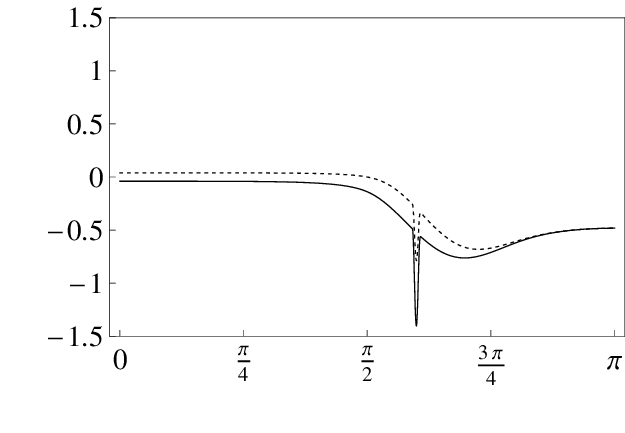} & 
\includegraphics[width=0.29\textwidth]{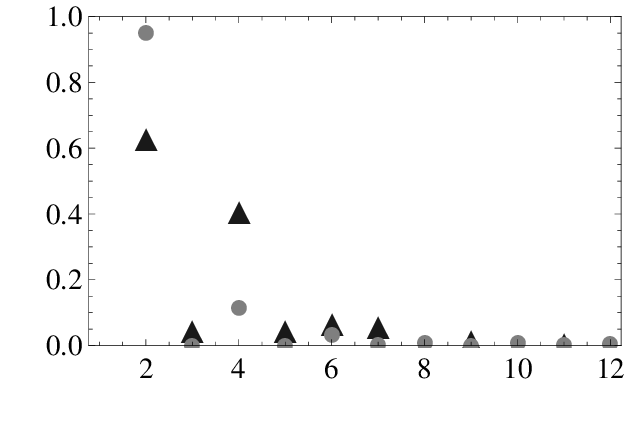} \\
\includegraphics[width=0.29\textwidth]{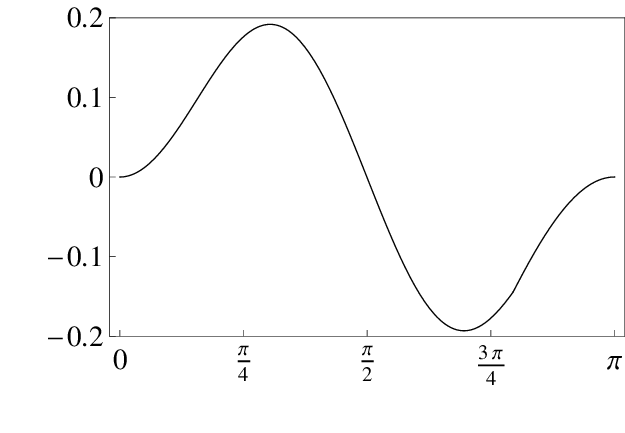} & 
\includegraphics[width=0.29\textwidth]{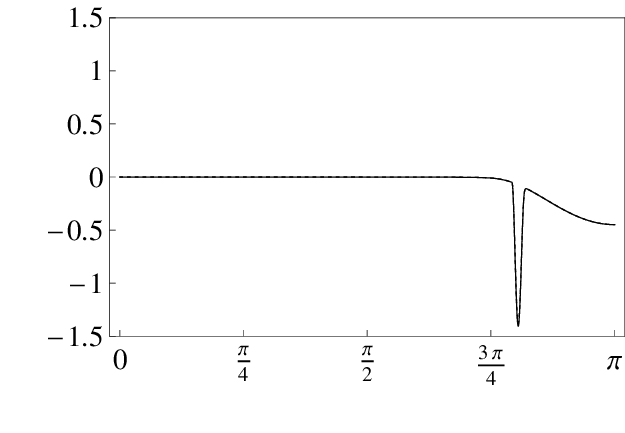} & 
\includegraphics[width=0.29\textwidth]{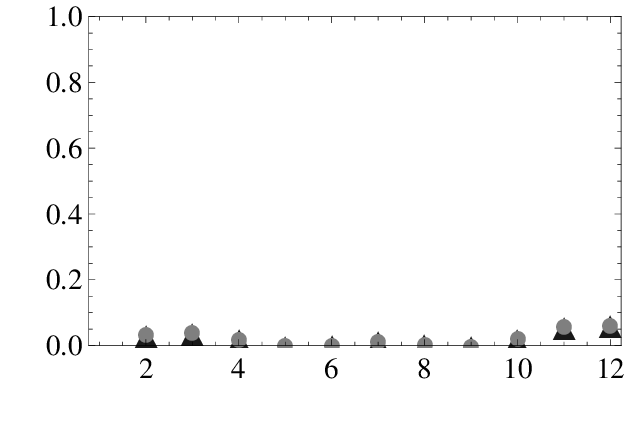}
\end{tabular}
\caption{\label{fig:SignalPlots}First column: rescaled temperature anisotropy $\tilde{\Theta}(\theta)\equiv \Theta(\theta,\phi)/[c_1\eta_{\rm obs}\cos(2\phi)]$ as a function of $\theta$. Second column: rescaled Stokes parameters $\tilde{Q}(\theta)\equiv 10^4\,Q(\theta,\phi)/[c_1\eta_{\rm obs}\cos(2\phi)]$ (solid curve) and $\tilde{U}(\theta)\equiv 10^4\,U(\theta,\phi)/[c_1\eta_{\rm obs}\sin(2\phi)]$ (dotted curve) as a function of $\theta$. Third column: rescaled $E$-mode $\tilde{C}^E_\ell\equiv 10^8[\ell(\ell+1)/2\pi]C^E_\ell/(c_1\eta_{\rm obs})$ (disks) and $B$-mode $\tilde{C}^B_\ell\equiv 10^8[\ell(\ell+1)/2\pi]C^B_\ell/(c_1\eta_{\rm obs})$ (triangles) power for small $\ell$.  From top to bottom, the rows correspond to $x_{\rm obs}/\eta_{\rm obs}=-0.8$, $-0.3$, $0$, $0.3$, $0.8$.}
\end{center}
\end{figure}

Our model contains only two free parameters:  $c_1$, which describes the amplitude of the perturbation, and $x_{\rm obs}$, which describes our $x$ coordinate relative to the incoming wave.  As explained in Section \ref{sec:EFT}, there is no temperature or polarization signal when $x_{\rm obs} < -\eta_{\rm obs}$, since in this situation we are outside of the causal future of the collision.  Meanwhile, from (\ref{hnot}) we see that $h_+$ is independent of time when $x_{\rm obs}\geq\eta_{\rm obs}$.  In this case 
\bea
\Theta(\theta,\phi) &=& \frac{c_1}{2}\left(\eta_{\rm obs}-\eta_{\rm rec}\right)
\cos(\theta)\sin^2(\theta)\cos(2\phi) \label{octa}\\
&=& \sqrt{\frac{2\pi}{105}}\,c_1\!\left(\eta_{\rm obs}-\eta_{\rm rec}\right)
\Big[Y_{32}(\theta,\phi)+Y^*_{32}(\theta,\phi)\Big] \,, \label{oct}
\eea
and therefore the CMB temperature anisotropy is a pure octopole.  In Figure \ref{fig:SignalPlots} we display the CMB temperature anisotropy for several values of $x_{\rm obs}$ between $-\eta_{\rm obs}$ and $\eta_{\rm obs}$.

An interesting feature of these plots is that the temperature anisotropy $\Theta$ is non-zero across the entire CMB sky.  To elaborate, note that for an observer located at $x=x_{\rm obs}$, the causal future of the collision at the time of last scattering subtends a maximum polar angle
\beq
\tc = \arccos\!\left( \frac{\eta_{\rm rec}+x_{\rm obs}}
{\eta_{\rm rec}-\eta_{\rm obs}}\right) 
\simeq \arccos\!\left( -\frac{x_{\rm obs}}{\eta_{\rm obs}}\right) \,,
\label{edge}
\eeq
where in the second expression we take $|x_{\rm obs}|$ to be of order $\eta_{\rm obs}$.  Therefore, one might have guessed that $\Theta$ should be non-zero only when $\theta\leq\tc$ (for $\phi=0$).  The reason this intuition fails is the CMB is not simply a snapshot of the universe at the time of last scattering.  In particular, the temperature anisotropy $\Theta$ is sourced by the change in the metric perturbation $h_{ij}$ between the emission of a CMB photon at recombination and its detection at the time of observation.  Since for all values of $x_{\rm obs}$ displayed in Figure \ref{fig:SignalPlots} the causal future of the collision includes the point of detection $x=x_{\rm obs}$, for all of these values of $x_{\rm obs}$ the CMB photons receive some redshift/blueshift due to $h_{ij}$.  Evidently, the boundary of the causal future of the collision simply generates a discontinuity in the first derivative of $\Theta$ at $\theta=\tc$.

\subsection{Polarization}
\label{ssec:polarization}

CMB polarization is produced by the Thomson scattering of CMB photons by free electrons. Generally speaking, the polarization of light scattered by a charged particle is determined by the quadrupole moments of the incident radiation.  It is conventionally parametrized by the Stokes parameters $Q$ and $U$, which in the context of CMB radiation can be expressed as the line-of-sight integrals \cite{Hu:1999vq}
\bea
(Q \pm iU)(\hat{n}) = \frac{\sqrt{6}}{10} \int dr \, e^{-\tau(r)} 
\frac{d\tau}{dr} \sum_m\, \Theta_{2m}(r\hat{n})\, _{\pm 2}Y_{2m}(\hat{n}) \,,
\label{QU}
\eea    
where the $_{\pm 2}Y_{2m}(\hat{n})$ are the spin-weighted spherical harmonics with $s=\pm 2$ (as in \cite{Goldberg:1966uu}), the $\Theta_{2m}(r\hat{n})$ are quadrupole moments of the ``CMB'' that would be observed by a free elec\-tron at the point 
\beq
\{\eta,\bfx\}|_{r\hat{n}}=\{\eta_{\rm obs}-r,\bfx_{\rm obs}+r\hat{n}\} \,,
\label{rn}
\eeq
and finally $\tau(r)$ denotes the optical depth out to a comoving distance $r$,
\beq
\tau(r) = \int_{\eta_{\rm obs}-r}^{\eta_{\rm obs}} a(\eta)\,d\eta\,
\sigma_T\, n_e(\eta) \,,
\eeq
where $n_e$ is the number density of free electrons and $\sigma_T$ is the Thomson cross section.

It is customary to express $n_e$ in terms of the ionization fraction $\chi$, i.e.
\beq
n_e(\eta) = \chi(\eta)\,n_p(\eta) = \chi(\eta)\,
\frac{a^3(\eta_{\rm obs})}{a^3(\eta)}\,n_p(\eta_{\rm obs}) \,,
\eeq
where $n_p$ is the number density of protons, excluding those in helium or any heavier elements.  The second expression notes that during the times of interest the number density of protons per comoving volume is constant (the effects of stellar burning are negligible).  Qualitatively speaking, the ionization fraction $\chi$ is near unity for times well before recombination, falls to near zero during recombination, remains near zero until re\-ion\-i\-za\-tion, and then abruptly rises back to near unity, remaining there until the present.  To model this evolution more precisely, we divide it into two regimes.  For times surrounding recombination, we use the RECFAST code \cite{Seager:1999bc}, inserting maximum-likelihood cosmological parameters from the WMAP seven-year data \cite{Komatsu:2010fb}.  For times surrounding reionization, we use the fitting function \cite{Lewis:2008wr}
\beq
\chi(z) = \frac{A}{2}\left\{1-\tanh\left[\frac{2}{3}
\frac{(1+z)^{3/2}-(1+z_{\rm re})^{3/2}}{(1+z_{\rm re})^{1/2}\Delta_z}
\right] \right\} , \qquad
z\equiv \frac{a(\eta_{\rm obs})}{a(\eta)} -1 \,,
\eeq       
with $A=1.08$, $z_{\rm re}=10.4$, and $\Delta_z=0.5$.  (The ionization fraction can exceed unity because the effects of helium ionization are included in $\chi$ but according to convention the protons in helium are not included in $n_p$.)  Finally, to model the scale-factor dependence of $\tau$ and $n_e$ as they appear in (\ref{QU}), we use the $\Lambda$CDM model with maximum-likelihood cosmological parameters from the WMAP seven-year data.

\begin{figure}[t!]
\begin{center}
\begin{tabular}{ccccc}
\includegraphics[width=0.23\textwidth]{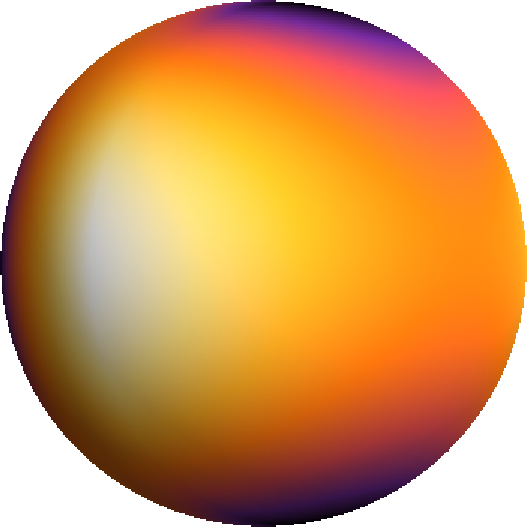} & \phantom{spi} &
\includegraphics[width=0.23\textwidth]{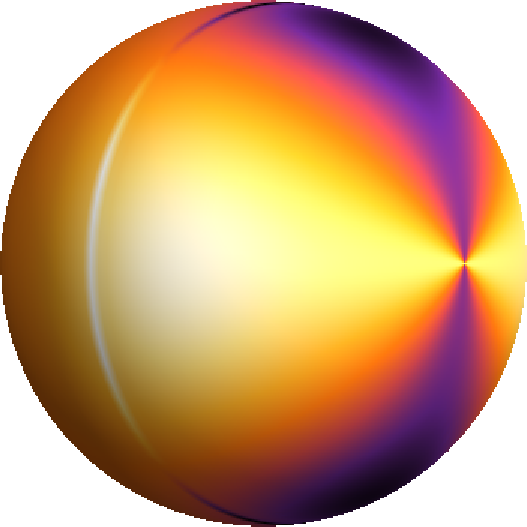} & \phantom{spi} &
\includegraphics[width=0.23\textwidth]{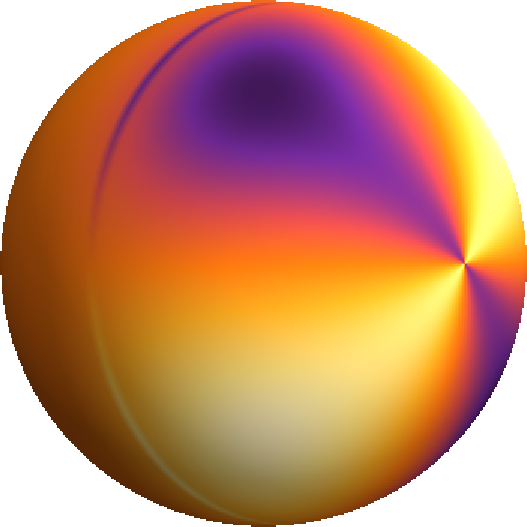} \\
& & & & \\
\end{tabular}
\caption{\label{fig:sky}Temperature anisotropy $\Theta$ (left panel), Stokes parameter $Q$ (center panel), and Stokes parameter $U$ (right panel) on 2-spheres representing our CMB sky, for $x_{\rm obs}=0.3\,\eta_{\rm obs}$.  Here darker colors correspond to more negative values, lighter colors to more positive values.}
\end{center}
\end{figure}

We are now prepared to compute the Stokes parameters $Q$ and $U$.  Recall that if $x_{\rm obs} < -\eta_{\rm obs}$, then these parameters must be zero by causality.  On the other hand, if $x_{\rm obs} > \eta_{\rm obs}$, then all ``observers'' along the null ray (\ref{rn}) see a pure octopole temperature anisotropy, i.e. the quadrupole component is precisely zero, and so $Q$ and $U$ are likewise zero.  For values of $x_{\rm obs}$ between $-\eta_{\rm obs}$ and $\eta_{\rm obs}$, we integrate (\ref{QU}) numerically.  Note that since the temperature anisotropy moments $\Theta_{\ell m}$ are non-zero only when $m=\pm 2$, the angular dependences of $Q$ and $U$ take the form $f(\theta)\cos(2\phi)$.  In Figure \ref{fig:SignalPlots} we display the Stokes parameters for several values of $x_{\rm obs}$ between $-\eta_{\rm obs}$ and $\eta_{\rm obs}$.  To further aid visualization, in Figure \ref{fig:sky} we display $\Theta$, $Q$, and $U$ on 2-spheres representing our CMB sky, for $x_{\rm obs}=0.3\,\eta_{\rm obs}$.

To understand these plots, recall that $Q$ and $U$ are computed by integrating from the point of observation to the surface of last scattering along a null ray with direction $\hat{n}$, with each point along the integration contributing to the integral according to the ``CMB'' that would be observed from that point.  For contributions to $Q$ and $U$ from very near the time of last scattering, the comoving size of the apparent horizon is relatively small, $\eta_{\rm rec}\ll\eta_{\rm obs}$, and so the aforementioned ``CMB'' probes a correspondingly small comoving scale.  On such small scales, any quadrupole of the temperature anisotropy $\Theta$ is invisible, except near the edge of the perturbation.  Meanwhile, the edge of the temperature anisotropy at the time of last scattering is located at the polar angle $\tc$ given by (\ref{edge}) (for $\phi=0$).  Thus, the Stokes parameters $Q$ and $U$ receive a contribution sharply peaked at $\theta=\tc$ from the temperature anisotropy at recombination.  Soon after recombination, the ionization fraction is near zero and as we integrate in from the surface of last scattering there is negligible contribution to $Q$ and $U$ until reionization.  By then, the comoving size of the apparent horizon is of order the present-day apparent horizon, and the edge of the temperature anisotropy extends to larger polar angles.  Thus, the Stokes parameters receive more broad contributions coming from CMB photon scattering after reionization.  This ``two-scale'' polarization signal is also seen with respect to fully 4D bubble collisions \cite{Czech:2010rg}, however those collisions do not produce gravity waves and therefore only generate $E$-mode polarization.

As alluded to above, CMB polarization can also be expressed in terms of $E$-mode and $B$-mode multipole moments.  These are defined according to 
\bea
(E_{\ell m} \pm i B_{\ell m}) = - \int d\Omega\, (Q \pm iU)(\hat{n})\, _{\pm 2}Y^*_{\ell m}(\hat{n}) \,.  
\eea
As with the temperature anisotropy multipoles $\Theta_{\ell m}$, only those moments with $m=\pm 2$ are non-zero. With our anisotropic bubble collisions, the $E_{\ell 2}$ are real and the $B_{\ell 2}$ are imaginary.  It is common to express CMB multipole moments such as $E_{\ell m}$ and $B_{\ell m}$ in terms of their total power $C_\ell$ at a given $\ell$, averaging over the moments $m$.  Accordingly we write
\beq
C^E_\ell \equiv \frac{1}{2\ell+1}\sum_m |E_{\ell m}|^2 \,,
\label{Cdef}
\eeq
and likewise for $B_{\ell m}$.  On the other hand, unlike with a statistically isotropic distribution such as the inflationary spectrum, with our signal the averaging over $m$ in $C^{E,B}_\ell$ discards non-trivial information.  Thus, an optimized search for anisotropic bubble collisions would make use of the full angular dependence of the polarization.  We leave the design of such a search to future work, and simple note the $C^E_\ell$ and $C^B_\ell$ as measures of the size of the signal.  These are displayed in Figure \ref{fig:SignalPlots} for several values of $x_{\rm obs}$ between $-\eta_{\rm obs}$ and $\eta_{\rm obs}$.  Note that the power in $E$-modes and in $B$-modes is comparable.

\section{Prospects for observation}
\label{sec:prospects}

The expected number of anisotropic bubble collisions for which the boundary of the causal future of the collision intersects the surface of last scattering is \cite{Salem:2010mi}   
\beq
{\cal N}_{\rm bub} = 8\pi\,\frac{\Hp}{\Hinf}\,\frac{\Gamma}{\Hp^3}\,\sqrt{\Omega_k}\,,
\label{N2b}
\eeq
where we have used $r_\star\equiv \eta_{\rm obs}-\eta_{\rm rec}\simeq \eta_{\rm obs}\simeq 2\sqrt{\Omega_k}$ to translate the corresponding expression in \cite{Salem:2010mi} into our notation (see Section \ref{ssec:4DFRW}).  As explained in Section \ref{sec:signal}, these collisions produce both a temperature anisotropy and a polarization anisotropy in the CMB, thus giving rise to distinctive ob\-ser\-va\-tion\-al signatures.  We call these ``partial-sky'' collisions.    

As shown in Section \ref{ssec:temp}, anisotropic bubble collisions produce a temperature an\-i\-so\-tro\-py even when the boundary of the causal future of the collision encloses the entire surface of last scattering.  For all but a negligible fraction of these collisions---the exceptions being those whose boundary passes within a comoving distance 2$\eta_{\rm rec}$ of the surface of last scattering---the temperature anisotropy is a pure octopole, and there is no associated polarization anisotropy.  Therefore, notwithstanding any distinctive ob\-serv\-able signatures that our analysis overlooks, the signal coming from these collisions is indistinguishable from the  octopole due to inflationary perturbations.  We call these ``full-sky'' collisions.  The expected number of them has not been presented in the literature, but it is straightforward to calculate in an\-al\-ogy to the 4D analysis of \cite{Salem:2011qz}.  The result is
\beq
{\cal N}_{\rm tot} = 2\pi\,\frac{\Hp}{\Hinf}\,\frac{\Gamma}{\Hp^3} 
= \frac{1}{4\sqrt{\Omega_k}}\,{\cal N}_{\rm bub}\,.
\label{N2c}
\eeq
Since the size of the CMB octopole has been measured, this result can be used to constrain the sizes of the other effects coming from anisotropic bubble collisions.

To better understand this constraint, note that each full-sky collision produces a temperature perturbation of the form (\ref{oct}), when expressed in terms of a spherical-polar coordinate system that places the center of the perturbation at $\theta=0$.  To express one such perturbation in a coordinate system that is centered on another, we simply rotate (\ref{oct}).  Since each perturbation corresponds to a pure octopole, the sum over a collection of rotated perturbations also corresponds to a pure octopole.  Meanwhile, the centers of these perturbations are distributed essentially uniformly in terms of the a\-zi\-mu\-thal angle of symmetry in the sky.\footnote{This is shown with respect to partial-sky collisions in \cite{Salem:2010mi}; it is straightforward to work in analogy to the 4D analysis of \cite{Salem:2011qz} (for example) to confirm that this applies to the centers of full-sky collisions as well.}  Therefore, summing over the perturbations produced by ${\cal N}_{\rm tot}$ full-sky collisions and taking the expectation value with respect to the various centers of these perturbations, we find the total power in the octopole to be
\bea
\< C^{\Theta}_{\ell=3} \>_{\rm rms}\approx \frac{\pi}{735}\frac{{\cal N}_{\rm bub}}{\sqrt{\Omega_k}}\,\big(
\<c_{1}\>^{\mbox{\scriptsize f-s}}_{\rm rms}\,\eta_{\rm obs}\big)^2\,,
\label{octoc}
\eea
where $C^\Theta_\ell$ is computed in analogy to $C^{E}_{\ell}$ in (\ref{Cdef}), the brackets and subscript rms note that we have taken the root-mean-square with respect to a random distribution of perturbations centers, $\<c_{1}\>_{\rm rms}^{\mbox{\scriptsize f-s}}$ denotes the root-mean-square of the full-sky collision amplitudes $c_1$, since these in general depend on the circumstances of a each collision, and we have used (\ref{N2c}) and $\eta_{\rm rec}\ll\eta_{\rm obs}$.  Comparing this to the size of the octopole anisotropy in the CMB (which is presumably dominated by the inflationary perturbation), we require \cite{Komatsu:2010fb}
\beq
C^\Theta_{\ell=3}\lesssim 10^{-10} \,.
\eeq

The severity of this constraint depends on the circumstances.  In particular, we find
\beq
c_1\eta_{\rm obs} \lesssim 
\left(\frac{c_1}{\<c_{1}\>_{\rm rms}^{\mbox{\scriptsize f-s}}}\right)\!
\left(\frac{\Omega_k}{10^{-5}}\right)^{\!\!1/4}\!
\left(\frac{3}{{\cal N}_{\rm bub}}\right)^{\!\!1/2}\! \times 5\times 10^{-6} \,.
\label{constraintF}
\eeq  
Thus, if the amplitude of a given partial-sky perturbation is the same as the amplitude of a typical full-sky perturbation, then under the favorable conditions $\Omega_k\approx 10^{-5}$ (see the introduction) and ${\cal N}_{\rm bub} = 3$ (for example), the amplitude of the perturbation is suppressed by about a factor of two relative to that of the inflationary temperature anisotropies.  Nevertheless, partial-sky anisotropic bubble collisions also produce $B$-mode polarization perturbations.  Therefore, even though the amplitude of the collision perturbation is somewhat suppressed relative to the amplitude of the inflationary tem\-per\-a\-ture an\-i\-so\-tro\-pies, it can still dominate over the inflationary $B$-mode polarization anisotropies, since the amplitude of the latter is unknown and could in principle be very small.
 
In Section \ref{sec:signal} we found that each partial-sky anisotropic bubble collision produces a polarization signal of order $[\ell(\ell+1)/2\pi]\,C^{E,B}_\ell\sim 10^{-8}\,(c_1\eta_{\rm obs})^2$ for low $\ell$ (the prefactor is included to conform to convention).  Combining this with the constraint (\ref{constraintF}), we find 
\beq
\frac{\ell(\ell+1)}{2\pi}\,C^{E,B}_\ell \lesssim 
\left(\frac{\<c_1\>_{\rm rms}^{\mbox{\scriptsize p-s}}}
{\<c_{1}\>_{\rm rms}^{\mbox{\scriptsize f-s}}}\right)^{\!2}\!
\left(\frac{\Omega_k}{10^{-5}}\right)^{\!\!1/2}\!
\times 8\times 10^{-19}\,,
\label{constraintG}
\eeq   
where $\<c_1\>_{\rm rms}^{\mbox{\scriptsize p-s}}$ is the root-mean-square amplitude for the partial-sky collisions.  As we briefly mentioned in the introduction, the gravitational lensing of the CMB due to intervening dark matter also generates a $B$-mode polarization signal.  This signal is of the same order as the upper bound in (\ref{constraintG}), if we set all of the parenthetical factors in (\ref{constraintG}) to unity \cite{Lewis:2006fu}.  However, the low-$\ell$ $B$-modes due to lensing can be subtracted out of the CMB to high precision using the lensing potential deduced from high-$\ell$ $E$-mode/$B$-mode correlations.  Indeed, it has been claimed that that the low-$\ell$ $B$-mode polarization power due to lensing can be reduced by at least a factor of 40 using these techniques \cite{Seljak:2003pn}.  If one is only interested in the lowest few $\ell$, where our signal is peaked, it is conceivable that even better lensing subtraction can be performed.  Finally, we stress that while we have focused on the statistics $C^{E,B}_\ell$, the polarization signals from anisotropic bubble collisions contain localized angular information, and this information correlates with a localized temperature anisotropy.  With an appropriate search algorithm, this should greatly enhance the possibility of detection.  

Under less favorable conditions than those considered above, $\Omega_k$ could be smaller and/or ${\cal N}_{\rm bub}$ could be larger.  Since $\Omega_k$ appears under a fourth root in (\ref{constraintF}) and under a square root in (\ref{constraintG}), the tightening of these constraints due to smaller values of $\Omega_k$ is somewhat mitigated.  (This statement treats the product $c_1\eta_{\rm obs}$ as a free parameter, meaning the decrease in $\eta_{\rm obs}\propto\sqrt{\Omega_k}$ due to any decrease in $\Omega_k$ is offset by contemplating the possibility that $c_1$ might be correspondingly larger.)  Larger ${\cal N}_{\rm obs}$ is not a problem for the overall power in the polarization signal, since the increased number of collisions offsets the stronger constraint on the amplitude of each collision implied by (\ref{constraintF}).  However, in terms of the distinctive effects of partial-sky collisions studied in Section \ref{sec:signal}, it is not clear that having more bubble collisions makes these effects easier to detect, so as to overcome the stronger constraint on their amplitude.  Note that ${\cal N}_{\rm bub}$ could be significantly less than unity, but we still observe a bubble collision out of fortuitousness.  On the other hand, the prediction for ${\cal N}_{\rm bub}$ involves a product of factors, one of which is very large, while the others are expected to be very small (see (\ref{N2b})).  It therefore seems most probable that this number is either very small or very large.

The situation where the number of partial-sky collisions is very large was studied in the context of a 4D parent vacuum in \cite{Kozaczuk:2012sx}, so as to determine whether the resulting spectrum of temperature perturbations could correspond to the nearly scale-invariant, Gaussian perturbations that are commonly attributed to the inflaton.  The conclusion was negative, due to an excess of power on large scales.  We expect this to also be the case with anisotropic bubble collisions, which would additionally fail to explain the observed temperature perturbations due to the anisotropic distribution of their effects across the CMB sky.

Before concluding, we remark on one final issue.  One might be concerned that, in the context of the globally eternally inflating spacetime that is implicit in the cosmology we are exploring, that all else being equal it is much more likely for our bubble to have nucleated in a 4D (or higher D) parent vacuum, as opposed to a 3D parent vacuum, because of the relative volume expansion factors.  In fact, in an eternally inflating spacetime all vacua are realized with diverging volume, and to properly perform this comparison requires defining a consistent regulator---called a measure---to manage these diverging volumes.  Although determining the ``correct'' measure is a major unresolved problem (for some recent reviews of this issue see \cite{Guth:2007ng,Freivogel:2011eg,Salem:2011qz}), the intuition garnered from phenomenologically successful measures is that the above reasoning is invalid.  In particular, measures that include such volume expansion factors violently disagree with observation, both in fully 4D cosmologies (here adopting some modest assumptions about the landscape) \cite{Feldstein:2005bm,Garriga:2005ee,Graesser:2006ft}, and in trans-di\-men\-sion\-al cosmologies such as we are exploring (without any assumptions about the landscape) \cite{SchwartzPerlov:2010ne}.

\section{Conclusions}
\label{sec:conclusion}

We have computed some of the observational signatures of anisotropic bubble collisions.  In the introduction we stressed several fortuitous circumstances that are necessary for these col\-li\-sions to occupy our past lightcone, while in Section \ref{sec:prospects} we discussed the conditions necessary for their effects to be discernible and yet not already ruled out.  At the end of Section \ref{ssec:toy} we made an assumption about the relative sizes of decay rates or, alternatively, the symmetries of instantons describing non-decompactifying decays.  While this might seem like a discouraging list of suppositions, at each point we have argued for the plausibility of this scenario.

Meanwhile, the observational signatures of anisotropic bubble collisions are very distinct.  Each collision produces a localized $E$-mode and $B$-mode CMB polarization perturbation, in addition to a temperature perturbation.  The profile of each of these perturbations features a $\cos(2\phi)$ rotation symmetry about the line-of-sight axis, due to $y$- and $z$-translation invariance in a Cartesian coordinate system with the collision perturbation incoming from the $x$ direction, and each perturbation profile is centered at a point along a great circle in the CMB sky, due to the $z$-translation invariance of the tunneling instanton describing the decompactification of the $z$ dimension.  It should be stressed that since the amplitude of the inflationary $B$-mode polarization signal is unknown and could in principle be very small, it is possible for the $B$-mode polarization effects of anisotropic bubble collisions to dominate over this background.  Note that these signatures contrast with those from bubble collisions in a 4D parent vacuum, as the latter do not produce significant $B$-mode polarization perturbations and are distributed essentially isotropically across the CMB sky.

Our analysis ignored the inflaton as a scalar degree of freedom in the early universe.  In a more realistic model, it is possible that the inflaton has some coupling with the metastable modulus that transmits the formation of our bubble and composes the bubble wall.  If so, then each anisotropic bubble collision would generate a scalar metric perturbation in addition to the tensor perturbation that we have computed, and this would modify the temperature and $E$-mode polarization perturbation produced by each collision.  However, this modification would simply add a radially symmetric profile to the above perturbations.  Note that any such coupling would not modify the $B$-mode polarization perturbation, since scalar inflaton perturbations cannot be a source of gravity waves.  

Future observations will probe the CMB with increasing sensitivity.  Although they might take a bit of luck to detect, the distinctive signatures of anisotropic bubble collisions would provide strong evidence for a spectacular extension of the standard cosmological model.

\acknowledgments

The authors thank Peter Graham and Surjeet Rajendran for collaboration during an early stage of this work.  For other helpful discussions, MPS thanks Jose Blanco-Pillado.  ES thanks Arpin\'e Davtyan for continued inspiration throughout the course of this project.  MPS and ES thank the organizers of the 2012 Cosmology and Complexity workshop in Hydra, and MPS also thanks the California Institute of Technology, for their warm hospitality. ES is supported by NSF grant PHY-0756174 and PS is supported by a Stanford Graduate Fellowship.  All of the authors are grateful for the support of the Stanford Institute for Theoretical Physics.

\providecommand{\href}[2]{#2}\begingroup\raggedright\endgroup

\end{document}